\documentclass[12pt]{article}

\usepackage{amssymb}
\usepackage{graphicx}
\renewcommand{\figurename}{Fig}

\usepackage{scicite}

\usepackage{times}

\newenvironment{sciabstract}{%
\begin{quote} \bf}
{\end{quote}}

\title{Quantum-critical scale invariance in a transition metal alloy}

\author
{Yasuyuki~Nakajima,$^{1,2\ast}$ Tristin~Metz,$^{2}$ Christopher~Eckberg,$^{2}$\\
Kevin~Kirshenbaum,$^{2}$ Alex~Hughes,$^{2}$ Renxiong~Wang,$^{2}$ Limin~Wang,$^{2}$\\
Shanta R.~Saha,$^{2}$ I-Lin Liu,$^{2,3,4}$ Nicholas P.~Butch,$^{2,4}$ Daniel~Campbell,$^{2}$\\
Yun Suk~Eo,$^{2}$ David~Graf,$^{5}$ Zhonghao~Liu,$^{6,7}$ Sergey V.~Borisenko,$^{6}$\\
Peter Y.~Zavalij,$^{8}$ Johnpierre~Paglione,$^{2,9\ast}$\\
\\
\normalsize{$^{1}$Department of Physics, University of Central Florida, Orlando, Florida 32816}\\
\normalsize{$^{2}$Maryland Quantum Materials Center, Department of Physics,}\\
\normalsize{University of Maryland, College Park, Maryland 20742}\\
\normalsize{$^{3}$Chemical Physics Department, University of Maryland, College Park, Maryland 20742}\\
\normalsize{$^{4}$NIST Center for Neutron Research, National Institute of Standards and Technology,}\\
\normalsize{Gaithersburg, Maryland 20899}\\
\normalsize{$^{5}$National High Magnetic Field Laboratory, Florida State University, Tallahassee, Florida 32310}\\
\normalsize{$^{6}$IFW-Dresden, Helmholtzstra{\ss}e 20, 01171 Dresden, Germany}\\
\normalsize{$^{7}$Shanghai Institute of Microsystem and Information Technology,}\\
\normalsize{Chinese Academy of Sciences, Shanghai 200050, China}\\
\normalsize{$^{8}$Department of Chemistry, University of Maryland, College Park, Maryland 20742}\\
\normalsize{$^{9}$The Canadian Institute for Advanced Research, Toronto, ON M5G 1Z8, Canada}\\
\\
\normalsize{$^\ast$To whom correspondence should be addressed;}\\
\normalsize{ E-mail: yasuyuki.nakajima@ucf.edu (Y.N.); paglione@umd.edu (J.P.).}
}

\date{}

\begin{document} 


\baselineskip24pt


\maketitle 

\def\23{Ba(Fe$_{1/3}$Co$_{1/3}$Ni$_{1/3}$)$_{2}$As$_{2}$}
\def\sr{Sr(Fe$_{1/3}$Co$_{1/3}$Ni$_{1/3}$)$_{2}$As$_{2}$}
\def\Co{BaCo$_{2}$As$_{2}$}
\def\Ni{BaNi$_{2}$As$_{2}$}
\def\Fe{BaFe$_{2}$As$_{2}$}
\def\FeCo{Ba(Fe,Co)$_{2}$As$_{2}$}
\def\tc{$T_{c}$}
\def\tn{$T_{N}$}
\def\hc2{$H_{c2}$}


\begin{sciabstract}
  Quantum-mechanical fluctuations between competing phases at $T=0$ induce exotic finite-temperature collective excitations that are not described by the standard Landau Fermi liquid framework \cite{hertz76,moriy85,milli93,colem05}. These excitations exhibit anomalous temperature dependences, or non-Fermi liquid behavior, in the transport and thermodynamic properties \cite{stewa01} in the vicinity of a quantum critical point, and are often intimately linked to the appearance of unconventional Cooper pairing as observed in strongly correlated systems including the high-$T_c$ cuprate and iron pnictide superconductors \cite{taill10,shiba14}. The presence of superconductivity, however, precludes direct access to the quantum critical point, and makes it difficult to assess the role of quantum-critical fluctuations in shaping anomalous finite-temperature physical properties. Here we report temperature-field scale invariance of non-Fermi liquid thermodynamic, transport, and Hall quantities in a non-superconducting iron-pnictide, {\23}, indicative of quantum criticality at zero temperature and zero applied magnetic field. Beyond a linear in temperature resistivity, the hallmark signature of strong quasiparticle scattering, we find the scattering rate that obeys a universal scaling relation between temperature and applied magnetic fields down to the lowest energy scales. Together with the dominance of hole-like carriers close to the zero-temperature and zero-field limits, the scale invariance, isotropic field response, and lack of applied pressure sensitivity suggests a unique quantum critical system that does not drive a pairing instability.
\end{sciabstract}

% In setting up this template for *Science* papers, we've used both
% the \section* command and the \paragraph* command for topical
% divisions.  Which you use will of course depend on the type of paper
% you're writing.  Review Articles tend to have displayed headings, for
% which \section* is more appropriate; Research Articles, when they have
% formal topical divisions at all, tend to signal them with bold text
% that runs into the paragraph, for which \paragraph* is the right
% choice.  Either way, use the asterisk (*) modifier, as shown, to

Non-Fermi liquid (NFL) behavior ubiquitously appears in iron-based high-temperature superconductors with a novel type of superconducting pairing symmetry driven by interband repulsion \cite{pagli10,shiba14}. The putative pairing mechanism is thought to be associated with the temperature-doping phase diagram, bearing striking resemblance to cuprate and heavy-fermion superconductors \cite{mazin08,kurok09a}. In iron-based superconductors, the superconducting phase appears to be centered around the point of suppression of antiferromagnetic (AFM) and orthorhombic structural order \cite{pagli10}. Close to the boundary between AFM order and superconductivity, the exotic metallic regime emerges in the normal state. The ``strange'' metallic behavior seems to be universal in strongly correlated metals near a quantum critical (QC) point, characterized by linear-in-$T$ resistivity \cite{hayes16,giral18,legro19,guo20}. The universal transport behavior is known as Planckian dissipation, where the transport scattering rate is constrained by thermal energy, $\hbar/\tau_{p}=k_{B}T$, where $\hbar$ is the reduced Planck constant and $k_{B}$ is the Boltzmann constant. Lacking an intrinsic energy scale, the scale-invariant transport in strange metals is one of the unresolved phenomena in condensed matter physics, but its microscopic origin has yet to be fully understood. In iron-based superconductors, along with the AFM order, the presence of an electronic nematic phase above the structural transition complicates the understanding of the SC and NFL behavior \cite{chu10,yi11,nakaj11a,chu12}. Moreover, the robust superconducting phase prohibits investigations of zero-temperature limit normal state physical properties associated with the QC instability due to the extremely high upper critical fields.

While AFM spin fluctuations are widely believed to provide the pairing glue in the iron-pnictides, other magnetic interactions are prevalent in closely related materials, such as the cobalt-based oxypnictides LaCoOX (X=P, As) \cite{yanag08}, which exhibit ferromagnetic (FM) orders, and Co-based intermetallic arsenides with coexisting FM and AFM spin correlations \cite{anand14,wieck15,wieck15a}. For instance, a strongly enhanced Wilson ratio $R_{W}$ of $\sim$ 7-10 at 2~K \cite{sefat09} and violation of the Koringa law \cite{anand14,wieck15,wieck15a} suggest proximity to a FM instability in {\Co}. {\Ni}, on the other hand, seems to be devoid of magnetic order \cite{sefat09a} and rather hosts other ordering instabilities in both structure and charge \cite{eckbe18}. Confirmed by extensive study, Fe, Co, and Ni have the same $2+$ oxidation state in the tetragonal ThCr$_2$Si$_2$ structure, thus adding one $d$ electron- (hole-) contribution by Ni (Fe) substitution for Co in {\Co} \cite{ni09,canfi09a,liu10,neupa11}, and thereby modifying the electronic structure subtly but significantly enough to tune in and out of different ground states and correlation types. Utilizing this balance, counter-doping a system to achieve the same nominal $d$ electron count as {\Co} can realize a unique route to the same nearly FM system while disrupting any specific spin correlation in the system.

Here, we utilize this approach to stabilize a novel ground state in the counter-doped non-superconducting iron pnictide {\23}, also nearly ferromagnetic but with a unique type of spin fluctuation that leads to very strong quasiparticle scattering. We show that NFL behavior is prevalent in the very low temperature charge transport and thermodynamic properties of {\23}, with temperature and magnetic energy scale invariance arising from a quantum critical ground state.

The hallmark of NFL behavior in {\23} is clearly observed in the resistivity (fig.~1a), which exhibits a quasi-linear $T$ dependence over three orders of magnitude variation, from 20~K down to at least 20~mK at $B=0$ T. In this temperature range, we find no discernible anomaly associated with phase transitions down to 20 mK, suggestive of the realization of an anomalous metallic ground state that persists to the $T=0$ limit. Furthermore, this behavior is strongly suppressed with magnetic field, which drives a recovery of Fermi liquid (FL) behavior (i.e., $\rho\propto T^2$) at low temperatures (See Supplementary Material (SM)).

%%%%%%  Mooij correlations and quantum interference
Note that the unusual resistivity observed in {\23} cannot be ascribed to either Mooij correlations \cite{ciuch18} or quantum interference \cite{lee85} due to randomness introduced by counter-doping. Given that the Mooij correlations are dominant, increasing randomness enhances the residual resistivity $\rho_0$, accompanied by a gradual change in the slope of $\rho(T)$ at high temperatures as observed LuRh$_{4}$B$_{4}$ \cite{lee85}. However, the overall slope of resistivity in {\23} is parallel shifted from that in {\Co} with a sizable increase of residual resistivity by $\sim 130~\mu\Omega$~cm, indicative of the absence of Mooij correlation (see SM). Also, the quasi-$T$-linear dependence of the resistivity at low temperatures in {\23} cannot be reproduced by the quantum corrections in conductivity caused by interference that provide the power law $\sigma \sim T^{p/2}$ (or $\rho \sim  T^{-p/2}$), where $p$ = 3/2 (dirty limit), 3 (electron-phonon scattering), or 1 (enhanced electron-electron interaction) \cite{lee85}. The absence of Mooij correlations and quantum interference allows us to treat scattering sources in charge transport independently. As demonstrated by a smooth change in the temperature slope of resistivity at $\sim$30 K (fig.~5a), the inelastic scattering dominates over the electron-phonon scattering in the charge transport at low temperatures.

%%%%%% 

Mimicking the quasi-linear behavior in the temperature dependence of $\rho(T)$ at 0~T, the magnetoresistance (MR) at 1.31~K $\Delta R(B)/R(0)$ varies sublinearly with applied field up to 35 T (fig.~1b). The quasi-linear-$T$ and $B$ dependence allow us to introduce a new energy scale involving the scattering rate, the quadrature sum of temperature and magnetic field $\Gamma(T,B)\equiv \sqrt{(k_{B}T)^{2}+(\eta \mu_{B}B)^{2}}$, where $\mu_{B}$ is the Bohr magneton and $\eta$ is a dimensionless parameter. Setting $\eta=0.67$, we find the unusual scaling in the inelastic scattering rate $\hbar/\tau=\hbar ne^{2}(\rho(T,B)-\rho(0,0))/m^{\ast}$, where $n$ is the carrier density extracted from low-temperature Hall coefficient measured at 0.5 T (fig.~3c) and $m^{\ast}$ is the effective mass obtained from low-temperature specific heat measured at 10 T (fig.~2b), as a function of $\Gamma (T,B)$, collapsing onto one universal curve as shown in fig.~1c, reminiscent of the observation in QC iron-pnictide BaFe$_{2}$(As,P)$_{2}$ \cite{hayes16}.

%%%%%%%%

The $\Gamma(T,B)$ scaling can closely be correlated to the Planckian bound of dissipation. Quantum mechanics allows the shortest dissipative time scale $\tau_{p}=\hbar/k_{B}T$, constrained by the uncertainty principle between dissipative time scale $\tau$ and energy dissipation $E\sim k_{B}T$, $\tau\cdot k_{B}T \gtrsim \hbar$. Redefining $\Gamma(T,B)$ as the dissipation energy scale in magnetic field, we can obtain the universal bound of dissipation, $\hbar/\tau_{p}\sim\Gamma(T,B)$. Our experimental observation in $\Gamma(T,B)$ scaling for the inelastic scattering gives a linear relation, $\hbar/\tau=A\Gamma(T,B)$ with $A=1.80$, in good agreement with expected behavior.

%%%%%

Notwithstanding the quasi-two-dimensional layered structure, the NFL magnetotransport is independent of applied field orientations with respect to the FeAs layers. We plot the anisotropy of the MR, $\Delta\rho(B\parallel c)/\Delta\rho(B\parallel ab)$, as a function of temperature in fig.~1d. The anisotropy between transverse MR in the out-of-plane field ($B\parallel c,~I\parallel ab$) and transverse MR in the in-plane field ($B\parallel ab,~B\perp I\parallel ab$) decreases down to unity with decreasing temperatures, suggesting the spatial dimension of the system is three. The isotropy in MR remains even at 35 T, as shown in the angular dependence of MR (fig.2 inset). Due to the three dimensionality, we observe similar $\Gamma(T,B)$ scaling in the resistivity regardless of applied field orientations (see SM). Moreover, the observed positive MR appears not to be driven by the orbital effect due to the Lorentz force, but rather associated with Zeeman energy-tuned scattering, evidenced by the isotropy in the MR between in-plane transverse ($B\parallel c,~I\parallel ab$) and longitudinal ($B\parallel I \parallel ab$) configurations (fig.1d).

In addition to resistivity, magnetic susceptibility $\chi=M/B$ and electronic heat capacity $C_{e}/T$ also exhibit canonical NFL behavior, i.e., diverging temperature dependence associated with QC instabilities \cite{stewa01}. The magnetic susceptibility varies as $\chi\propto T^{-1/3}$ at low temperatures below 8~K (inset of fig.~2a), in contrast to the $T$-independent Pauli paramagnetic susceptibility $\chi_{p}=2g\mu_B^2 D(E_F)$ (with electron $g$-factor and density of states at the Fermi energy $D(E_F)$) observed in FL metals, and observed upon increasing magnetic field to 7~T (fig.2a).
A similar crossover is also observed in the heat capacity. Obtained form the subtraction of phonon ($C_{ph}$) and nuclear Schottky contributions ($C_{Sch}$) from the total heat capacity ($C_{tot}$), the electronic specific heat coefficient $C_{e}/T = (C_{tot}-C_{ph}-C_{Sch})/T$ exhibits power law divergence, $C_{e}/T\sim T^{-0.25}$, stronger than logarithmic in the temperature dependence down to $\sim$150~mK (fig.2b). Diminished with applying field, the NFL behavior observed in zero field completely disappears at applied field of 10~T, indicative of the recovery of FL. We note that the obtained specific heat coefficient $\gamma = C_{e}/T$ at $B$ = 0 T, combined with the magnetic susceptibility $\chi$, provides large Wilson ratio $R_{W}=\pi^{2}k_{B}^{2}\chi/3\mu_{B}^{2}\gamma = 3.2$ at $T =$ 1.8 K, suggestive of the presence of magnetic instabilities similar to {\Co}.

The observation of FL recovery with magnetic field corroborates the presence of a new energy scale $k_B T^{\ast}$, distinctive of crossover between the QC $(k_B T \gg g\mu_B B)$ and FL $(k_B T \ll g\mu_B B)$ regimes. Intriguingly, this new energy scale allows a single scaling function of $T/B$ in the magnetization, written by,  
\begin{equation}
	-\frac{dM}{dT}=B^{-\frac{1}{3}}f_M\left (\frac{T}{B}\right ),
\end{equation}
as shown in fig.2c. This scaling relation indeed reveals the underlying free energy given by a universal function of $T/B$, 
\begin{equation}
	F(T,B)=B^{(d+z)/y_{b}}f_F\left(\frac{T}{B^{z/y_{b}}}\right),
\end{equation}
where $d$ is the spatial dimensionality, $z$ is the dynamic exponent, and $y_{b}$ is the scaling exponent related to the tuning parameter $B$ \cite{hertz76,moriy85,milli93,sachd99}. Here, $f_{F}(x)$ is a universal function of $x$. Hence, the magnetization can be derived from the derivative of the free energy,
\begin{equation}
	-\frac{dM}{dT}=-\frac{d}{dT}\left(\frac{dF}{dB}\right)=B^{d/y_{b}-1}f_M\left(\frac{T}{B^{z/y_{b}}}\right).
\end{equation}
Directly comparing this with the observed QC scaling relation in fig.~2c, we can extract the critical exponents in the free energy, namely, $z/y_{b}=1$ and $d/y_{b}-1=-1/3$, yielding $z = y_{b}$ and $d/z = 2/3$. These values of the critical exponents describe the specific heat by using the same free energy,
\begin{equation}
\frac{C_{e}(B,T)}{T}= -\frac{\partial^2F}{\partial T^2}=B^{\frac{d-z}{y_b}}f_{C}\left(\frac{T}{B^{z/y_{b}}}\right).
\end{equation}
Rewriting the free energy, $F(T,B)=B^{\frac{d+z}{y_b}}f(T/B^{z/y_{b}})=T^{\frac{d+z}{z}}\tilde{f}(B/T^{y_{b}/z})$, we find
\begin{equation}
  \frac{\Delta C_{e}(T,B)}{T} = \frac{C_{e}(T,B)}{T}-\frac{C_{e}(T,0)}{T}=B^{-\frac{1}{3}}g_{C}\left(\frac{T}{B}\right),
\end{equation}
where $g_{C}(x)$ is field-dependent part of $f_{C}(x)$ (see SM). As demonstrated in fig.~2d, this expression illustrates scale invariance in the specific heat that persists over nearly three orders of magnitude in the scaling variable $T/B$.

%Hall effect
The $T/B$ scaling in thermodynamics clearly discloses the presence of the QCP located exactly at zero field and absolute zero, similar to the layered QC metals YbAlB$_4$ \cite{matsu11} and YFe$_2$Al$_{10}$ \cite{wu14}. More notably, the multi-band nature in iron pnictides affixes the uniqueness of quantum criticality for {\23}. Dominated by electron-like carriers, the Hall resistivity $\rho_{yx}$ is negative and perfectly linear in field at high temperatures ($T=20$ K) as shown in fig.3a. Upon cooling, $\rho_{yx}$ develops a non-linearity with negative curvature. More prominent below 1~K, the non-linear Hall resistivity switches its sign at low fields below 2~T. The sign change is more readily observed in the temperature dependence of Hall coefficient $R_{H}$ defined by $\rho_{yx}/B$ at low-$T$ and low-$H$ region (fig.3b), implying that hole-like carriers dominate the transport in the vicinity of the QCP. The radial shape of the dominant carrier crossover in the field-temperature phase diagram confirms the absence of an intrinsic energy scale in $R_{H}$ (fig.~3c), or in other words, the presence of scale invariance in the Hall effect tuned by temperature and magnetic field. Similar to the resistivity, $R_{H}$ obeys $\Gamma(T,B)$ scaling (fig.~3d), consolidating the existence of scale invariance near the quantum critical point in this system beyond any doubt.

Angle-resolved photoemission measurements identify a unique electronic structure and confirm the anomalous scattering rate correlated with Planckian dissipation. Unlike heavily electron-doped BaCo$_{2}$As$_{2}$, the electronic structure for {\23} consists of a large hole-like pocket and a cross-shaped electron-like Fermi surface around the $\Gamma$ point together with oval and elongated electron pockets around the $M$ points, exhibited by the Fermi surface map (fig.4a), the band dispersion along $k_{x}=0$ direction (fig.~4b) at 30K, and a schematic illustration (fig.~4a, inset). The elongated electron pockets are very shallow, and the chemical potential is located close to the bottom of the shallow bands. Dominating transport at low temperatures and fields, the large hole-like pocket is identified as the one responsible for quantum critical behavior. Amazingly, the scattering rate (obtained from the dispersion of the hole-like bands at 1~K) varies linearly with the kinetic energy up to 100 meV, consistent with Planckian dissipation as observed in the resistivity (fig.~4c and d).

While our primary observations of the scale invariance in the thermodynamics are consistent with quantum criticality overall,
they indicate a highly unusual critical behavior in {\23}. While sharing an enhancement of the Wilson ratio with {\Co} and $T^{5/3}$ transport scattering rate indicative of a FM instability, the critical behavior in \\{\23} is not described by any known theoretical predictions. Assuming spacial dimensionality of three ($d=3$) based on the observed isotropic response in MR and magnetization (see SM), the observed critical exponents of $d/z=2/3$ and $z= y_b$ yield $z=y_{b}=4.5$.

% Utilizing the Josephson's identity $d\nu=2-\alpha$ \cite{golde92}, where $\nu$ and $\alpha$ are the correlation length exponent and the critical exponent of heat capacity ($C\sim T^{-\alpha}$), respectively, we can obtain $d\nu=2.75$, or $\nu = 0.92$, satisfying the Harris's criterion, $d\nu>2$, that describes the stability of critical behavior against point randomness \cite{harri74}.

The extracted dynamical exponents from our measurements does not match the predictions for either mean-field Hertz-Moriya-Millis theory for $d=3$ (which predict $z=3$ for clean FM and $z=4$ for dirty FM quantum criticality with $\nu=1/2$) \cite{hertz76,moriy85,milli93}, or predictions for clean FM beyond mean-field, which predict the appearance of a weak first-order transition with $z=3$ and $\nu=1/4$ for $d=3$ and quantum-wing critical points with the same critical exponents as the mean-field theory \cite{belit97,rech06,condu09,kirkp15,brand16}. Quantum critical behavior in disordered 3d FM has been well explained by the Belitz-Kirkpatrick-Vojta theory, predicting critical exponents $\nu =1$ and $z=3$ for the asymptotic limit and $\nu =0.25$ and $z=6$ for the preasymptotic limit \cite{brand16,sales17}, neither of which is in agreement with our observation. Experimentally, previously measured exponents in QC materials, such as YbNi$_{4}$(P$_{1-x}$As$_{x}$)$_{2}$ (FM QCP, $\nu z\sim5$) \cite{stepp13}, CeCu$_{6-x}$Au$_{x}$ (AFM QCP, $d/z=1/4,\nu z=1$) \cite{schro00}, $\beta$-YbAlB$_{4}$ (mixed-valence meal, $d/z=1/2,~\nu z =1$) \cite{matsu11}, YFe$_{2}$Al$_{10}$ (layered QC metal, $d/z=1,~\nu z =0.59$ \cite{wu14}), and Sr$_{0.3}$Ca$_{0.7}$RuO$_{3}$ (disordered FM QCP, $z=1.76$) \cite{huang15a} are also incompatible with the measured dynamical exponent.

%%%%%% Griffiths phase

The high residual resistivity observed in {\23} evokes the possible realization of quantum Griffiths phase where the quantum critical behavior is dominated by ferromagnetic rare regions.% However, our single-crystal x-ray refinement confirms that the counter-dopants into Co sites are randomly distributed and form a perfectly crystalline alloy without any clusters, inconsistent with the quantum Griffith phase picture (See SM).
The quantum Griffiths model predicts power-law singularities in the magnetic susceptibility ($\chi\sim T^{\lambda-1}$), specific heat ($C/T\sim T^{\lambda-1}$), and magnetization ($M\sim B^{\lambda}$), determined by the nonuniversal Griffiths exponent $\lambda$ that takes $0$ at the quantum critical point and increases with distance from criticality \cite{vojta10}. In the present system, however, $\lambda = 2/3$ extracted from the magnetic susceptibility ($\chi\sim T^{-1/3}$ (fig.1a inset))) disagrees with $\lambda = 0.75$ obtained from the specific heat ($C/T\sim T^{-0.25}$ (fig.2b inset)), irreconcilable with the quantum Griffiths model. Besides, the critical exponents in {\23} do not agree with those obtained experimentally in other quantum Griffith systems \cite{brand16}. For instance, disordered weak ferromagnet Ni$_{1-x}$V$_{x}$ show critical behavior dominated by quantum Griffiths singularities, $\chi\sim T^{\lambda-1}$ and $M\sim B^{\lambda}$, over a wide range of vanadium concentration \cite{ubaid10,wang17c}. On the other hand, in {\23}, $\lambda =2/3$ derived from the magnetic susceptibility contradicts $\lambda$ obtained from magnetization $M\sim B^{0.75}$ (See SM), in conflict with the quantum Griffiths phase.
%%%%%%%%%%%%%%%%%%%% 

Highly unusual dynamical critical behavior in this material cannot be simply explained by existing FM QCP theories, but instead, it can be attributed to substitutional alloying by counter-doping. Indeed, the anomalous behavior observed in {\23} is more prominent than that observed in both of the end members of the $3d^{7}$ configuration line, namely, {\Co} and Ba(Fe,Ni)$_{2}$As$_{2}$ (see SM), signifying that the specific 1/3 equal ratios of Fe:Co:Ni in {\Co} are indeed important to stabilizing a unique quantum critical ground state. In fact, as shown in fig.~5, the observed NFL scattering behavior in {\23} is completely robust against pressure and even replacement of Ba for Sr (i.e. in {\sr}), implying either an electronic structure modification beyond $d$-electron tuning, or a significant role for transition metal site dilution. In fact, while generally obscuring the critical behavior, high randomness due to substitution indeed plays an important role in some quantum critical materials, such as medium entropy alloys \cite{sales16,sales17}, in which similar NFL behavior has been realized \cite{sales16,sales17}. Together with the pressure insensitivity of the $T$-linear scattering in {\23}, our experimental observations of scale invariance in this system indicates that substitutional alloying is a key ingredient to tune the quantum criticality that may provide the key to understanding the lack of superconductivity driven by quantum critical fluctuations.

\bibliographystyle{Science}

\begin{figure*}[t]
\includegraphics[width=16cm]{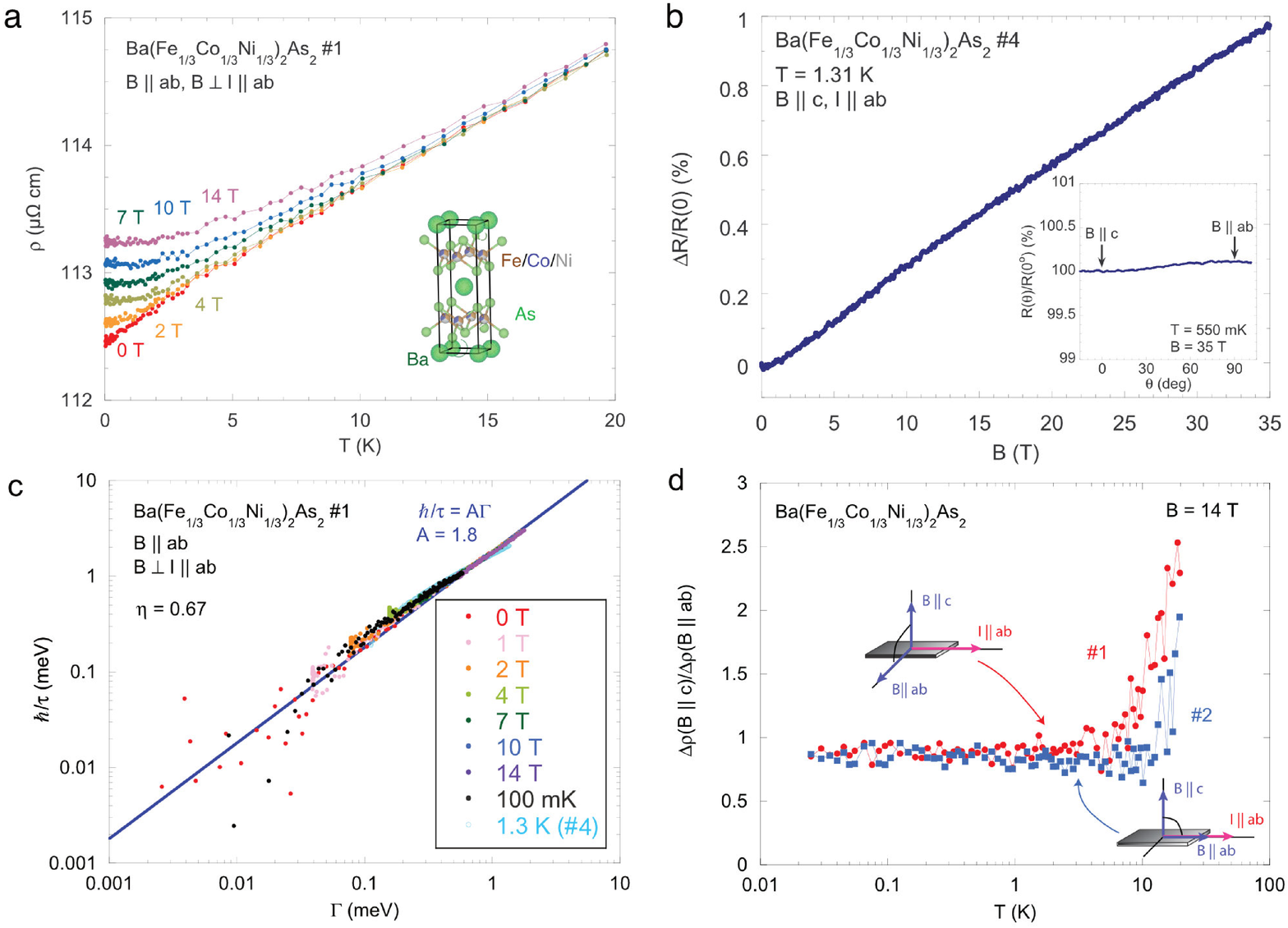}
\caption{{\bf Scale invariance in the resistivity of {\23}.}  {\bf a}, Temperture dependence of resistivity for \23 in the configuration of $B \parallel ab, B\perp I\parallel ab$. Inset: Crystal structure for {\23} \cite{momma11}. {\bf b}, Magnetic field dependence of $\Delta R(B)/R(0)\equiv (R(1.31~\mathrm{K},B)-R(1.31~\mathrm{K},0))/R(1.31~\mathrm{K},0)$ at $T$ = 1.31 K. Inset: Angular dependence of magnetoresistance at $T$ = 550 mK and $B (\parallel c \perp I)$ = 35 T. {\bf c}, Inelastic scattering rate $\hbar/\tau$ as a function of $\Gamma=\sqrt{(k_{B}T)^{2}+(\eta \mu_{B}B)^{2}}$ with $\eta =0.67$, suggestive of a universal scale invariance in the scattering mechanism in {\23}. A blue sold line is a linear fit to data using $\hbar /\tau =A\Gamma$ with $A$ = 1.8. {\bf d}, Temperature dependence of anisotropy of magnetoresistance between $\Delta\rho(B\parallel c)$ and $\Delta\rho(B\parallel ab \perp I)$ (sample \#1) and between $\Delta\rho(B\parallel c)$ and $\Delta\rho(B\parallel ab \parallel I)$ (sample \#2) at $B=14$ T, showing lack of anisotropy in the scattering rate.}
\end{figure*}

\begin{figure*}[t]
\includegraphics[width=15cm]{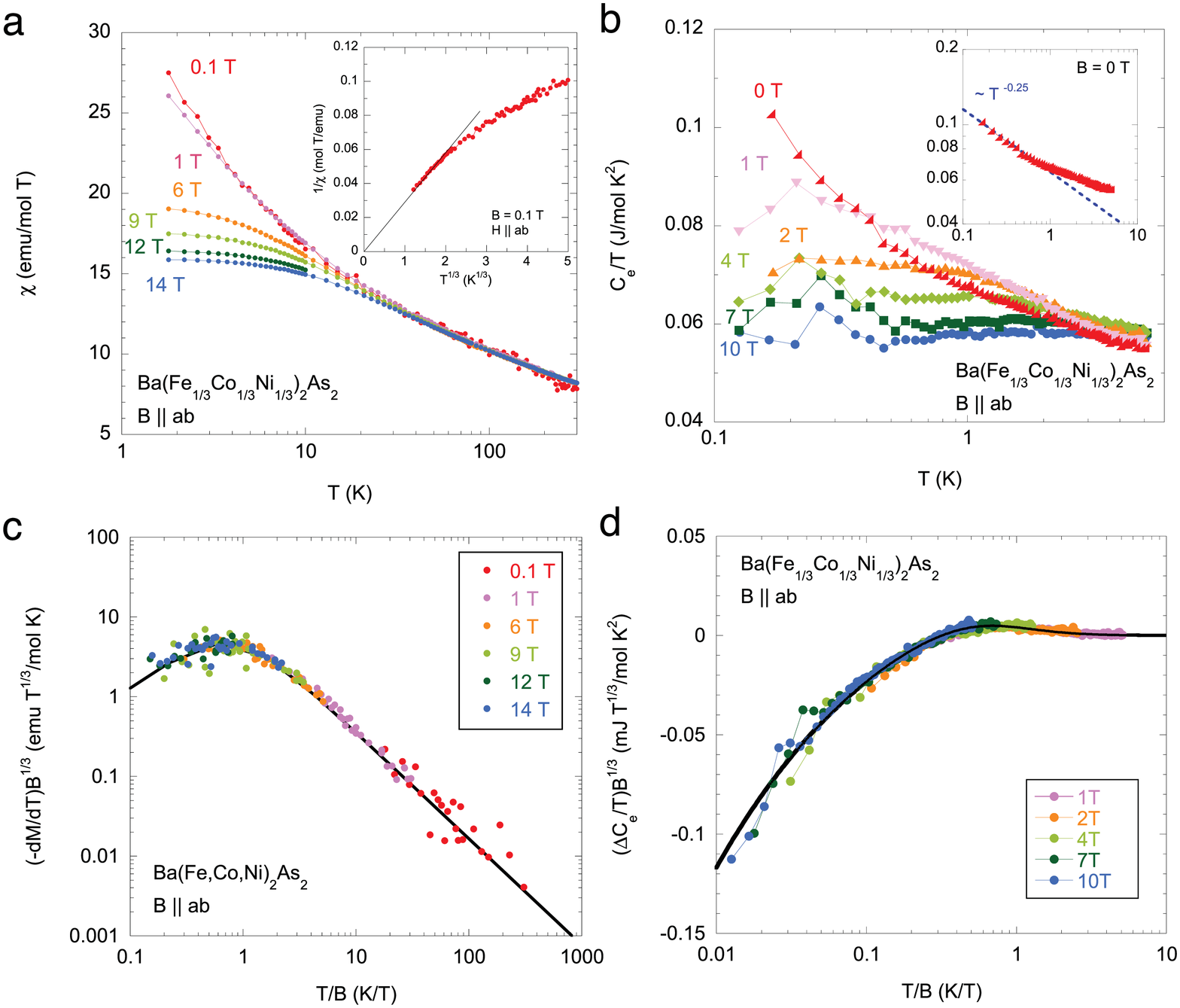}
\caption{{\bf Non-Fermi liquid to Fermi liquid crossover and scale invariance in thermodynamic quantities.} {\bf a}, Temperature dependence of magnetic susceptibility for {\23} ($B \parallel ab$). Inset: $1/\chi$ as a function of $T^{1/3}$ at low temperatures. {\bf b}, Electronic specific heat $C_{e}/T = (C_{tot}-C_{sch}-C_{lat})/T$ for {\23} under several fields parallel to $ab$-plane. Inset: log-log plot for $C_{e}/T$ vs $T$ for $B=0$ T. The dashed line emphasizes the $T^{-0.25}$-power law behavior observed below ~ 1 K. {\bf c}, Temperature-magnetic field scale invariance in magnetization and {\bf d}, specific heat. The measured magnetization and specific heat collapse onto universal scaling curves in the forms of $-dM/dT= B^{-1/3}f_{M}(T/B)$ and $\Delta C_e/T=B^{-1/3}g_{C}(T/B)$, respectively, indicating the presence of the underlying free energy given by a universal function of $T/B$ and the existence of the QCP located at $T=0$ and $B = 0$. Black lines represent scaling functions $f_{M}(T/B)$ for magnetization and $g_{C}(T/B)$ for specific heat derived from the underlying free energy, described in the main text.}
\end{figure*}

\begin{figure*}[t]
\includegraphics[width=16cm]{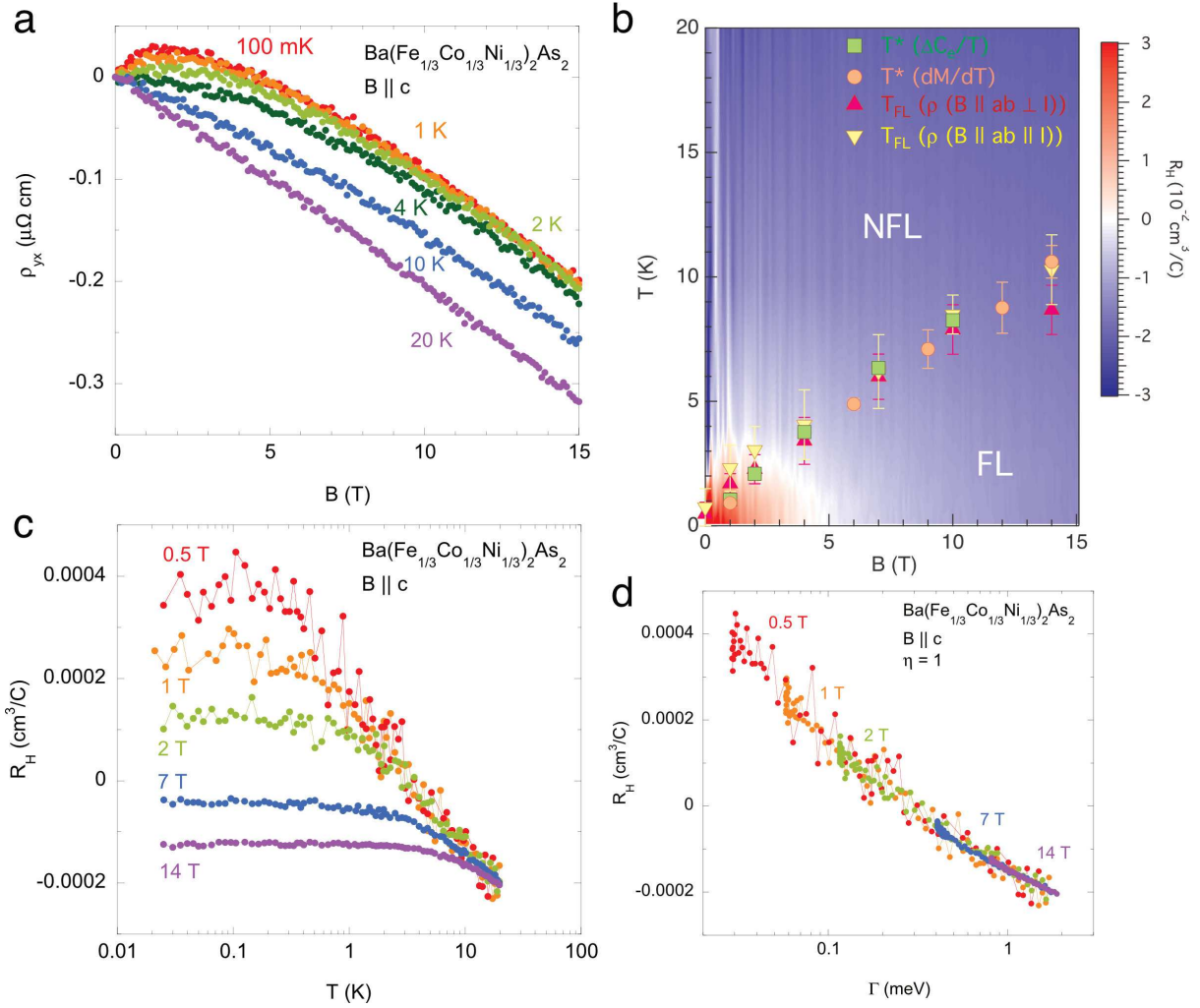}
\caption{{\bf Sign change due to dominant hole-like carriers near the quantum critical point and scale invariance in Hall effect.} {\bf a}, Hall resistivity $\rho_{yx}$ as a function of $B$. At high temperatures, $\rho_{yx}$ is negative and linear in field. Upon cooling temperatures, $\rho_{yx}$ becomes non-linear and its sign switches to positive at low fields below 2~T. {\bf b}, Temperature dependence of Hall coefficient $R_{H}$ defined by $\rho_{yx}/B$. {\bf c}, $T-B$ phase diagram with color plot of $R_{H}$. Crossover temperatures $T^{\ast}$ obtained from the quantum scaling in $dM/dT$ and $\Delta C_{e}/T$ and $T_{FL}$ from the $T^{2}$-fit are also plotted in the phase diagram. {\bf d}, $R_{H}$ as a function of $\Gamma(T,B) \equiv\sqrt{(k_{B}T)^{2}+(\eta \mu_{B}B)^{2}}$ with $\eta = 1$. All the data collapse onto one universal curve, suggesting unusual scaling between temperature and applied field similar to that found in the longitudinal resistivity.}
\end{figure*}

\begin{figure*}[htb]
  \includegraphics[width=17cm]{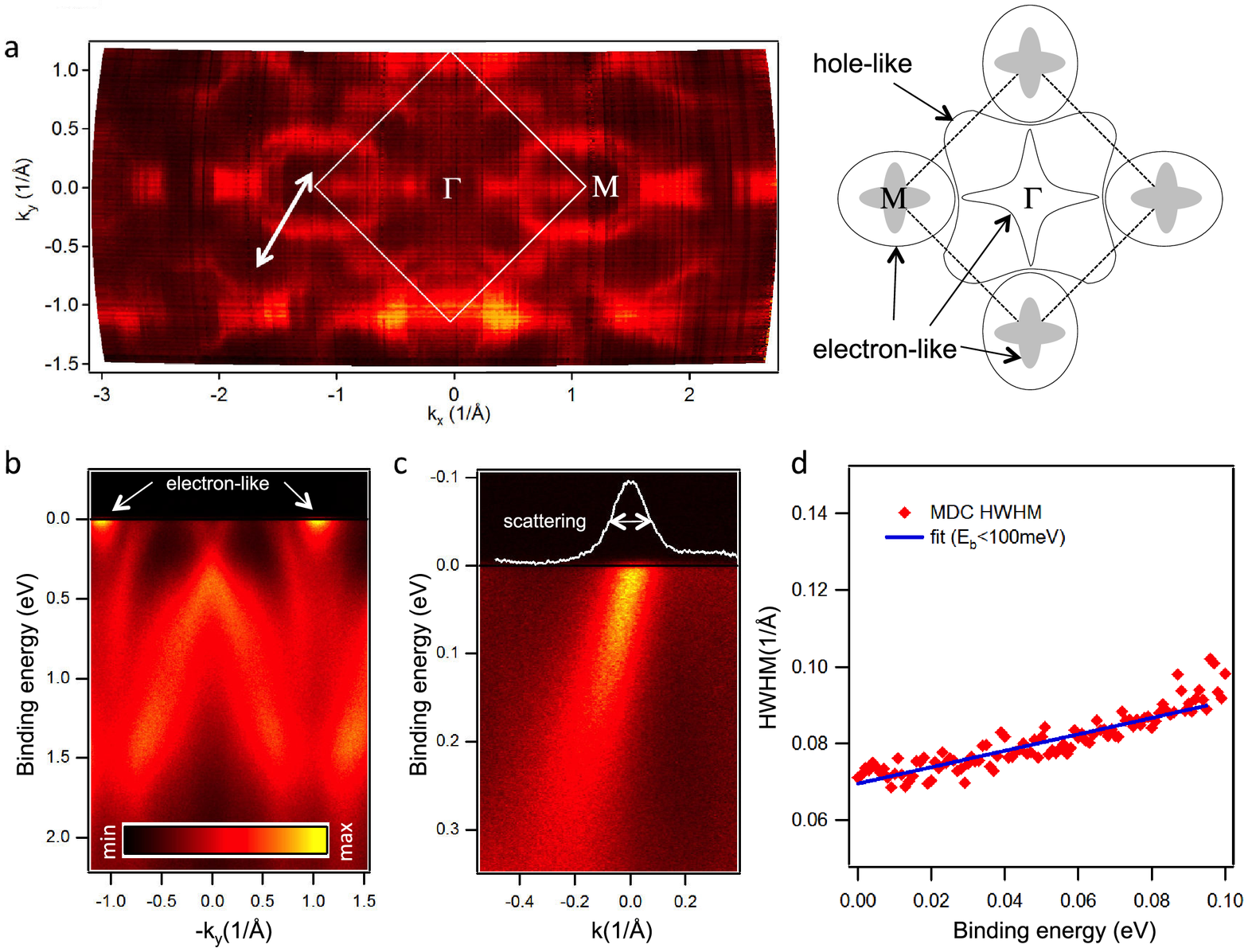}
  \caption{{\bf Fermi surfaces and anomalous scattering rates in {\23}.} {\bf a}, Angle-resolved photoemission study of Fermi surface map for {\23}, measured at 30 K. White lines denote the Brillouin zone (BZ) boundary, and white arrow corresponds to the cut shown in panel {\bf c}. The inset depicts a schematic Fermi surface corresponding to the experimentally observed data, with shallow elongated electron pockets shown in gray. {\bf b}, Energy cut along the $k_{x}=0$ direction, highlighting the elongated shallow electron-like pockets observed near the BZ corners. {\bf c}, Energy dispersion of hole-like pocket measured at 1 K near the BZ boundary along the cut indicated by white arrow in panel {\bf a}, where sharp crossings of the Fermi level are found. The momentum axis originates at the crossing point. The white spectrum is the momentum distribution curve at the Fermi level, with indicated width a representative measure of the scattering rate. {\bf d}, Scattering rate energy dependence obtained from the width of energy dispersion in panel {\bf c}, with linear fit up to 100 meV.}
\end{figure*}

\begin{figure*}[t]
\includegraphics[width=16cm]{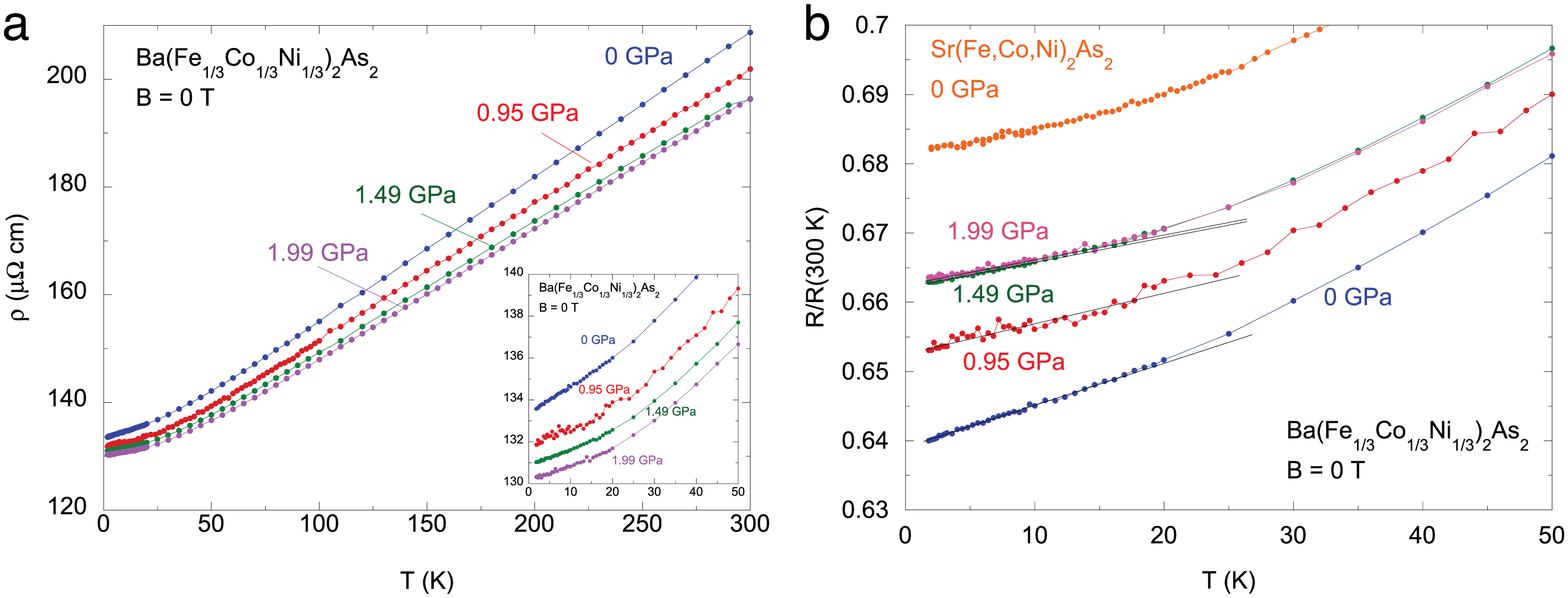}
\caption{{\bf Robust non-Fermi liquid behavior against pressure.} {\bf a}, Overall temperature dependence of resistivity for {\23} under applied pressure. The linear-$T$ resistivity below 20 K is robust against applying pressure up to 1.99 GPa as shown in the inset. {\bf b}, Normalized resistance $R/R(\mathrm{300~K})$ vs $T$ under pressure. On applying pressure, $R/R(\mathrm{300~K})$ for {\23} increases, approaching that for {\sr} with smaller lattice constants than {\23} (see SM), indicative of robustness of linear-$T$ behavior in the resistivity against pressure.}
\end{figure*}

\section*{Acknowledgments}
Experimental research was supported by the National Science Foundation Division of Materials Research Award No. DMR-1610349, and materials development supported by the Gordon and Betty Moore Foundation's EPiQS Initiative through grant no. GBMF4419. A portion of this work was performed at the National High Magnetic Field Laboratory, which is supported by National Science Foundation Cooperative Agreement No. DMR-1644779 and the State of Florida.
\section*{Supplementary materials}
Materials and Methods\\
Supplementary Text\\
Figs. S1 to S10\\
References \textit{(1-5)}

\clearpage

\begin{center}
\textbf{\large Supplementary Materials: Planckian dissipation and scale invariance in a quantum-critical disordered pnictide}
\end{center}

% supp. section modifications:
\newcommand{\beginsupplement}{%
        \setcounter{figure}{0}
        }
     
\renewcommand{\thesection}{S.\arabic{section}}
\renewcommand{\thesubsection}{\thesection.\arabic{subsection}}
\makeatletter
\renewcommand{\fnum@figure}
{\figurename~S\thefigure}
\makeatother
\renewcommand{\figurename}{FIG.}
\renewcommand{\theequation}{S\arabic{equation}}
\beginsupplement

\section{Growth and characterization}
The samples of {\23} were grown by {\it TM}As ({\it TM} = Fe, Co, and Ni) self-flux method with the molar ratios of 3:4:4:4 = Ba:FeAs:CoAs:NiAs. Resulting crystals were cleaved out of the flux. The typical crystal size is $5\times5\times0.1$ mm$^3$. 

% lattice constant
The lattice constants $a$ and $c$ for {\23} were determined by x-ray diffraction with Cu-$K_{\alpha}$ radiation, plotted together with  $ATM_{2}$As$_{2}$ ($A$=Sr and Ba, $TM$=Fe, Co and Ni) as a function of $d$ configurations (fig.S1). Sharing the same $d$ configuration with each other, the lattice parameter $c$ for {\23} is similar to those for BaCo$_{2}$As$_{2}$ and Ba(Fe$_{0.5}$Ni$_{0.5}$)$_{2}$As$_{2}$, while $a$ has a large variation by 1\% among them.

\begin{figure}[tbh]
\begin{center}
\includegraphics[width=12cm]{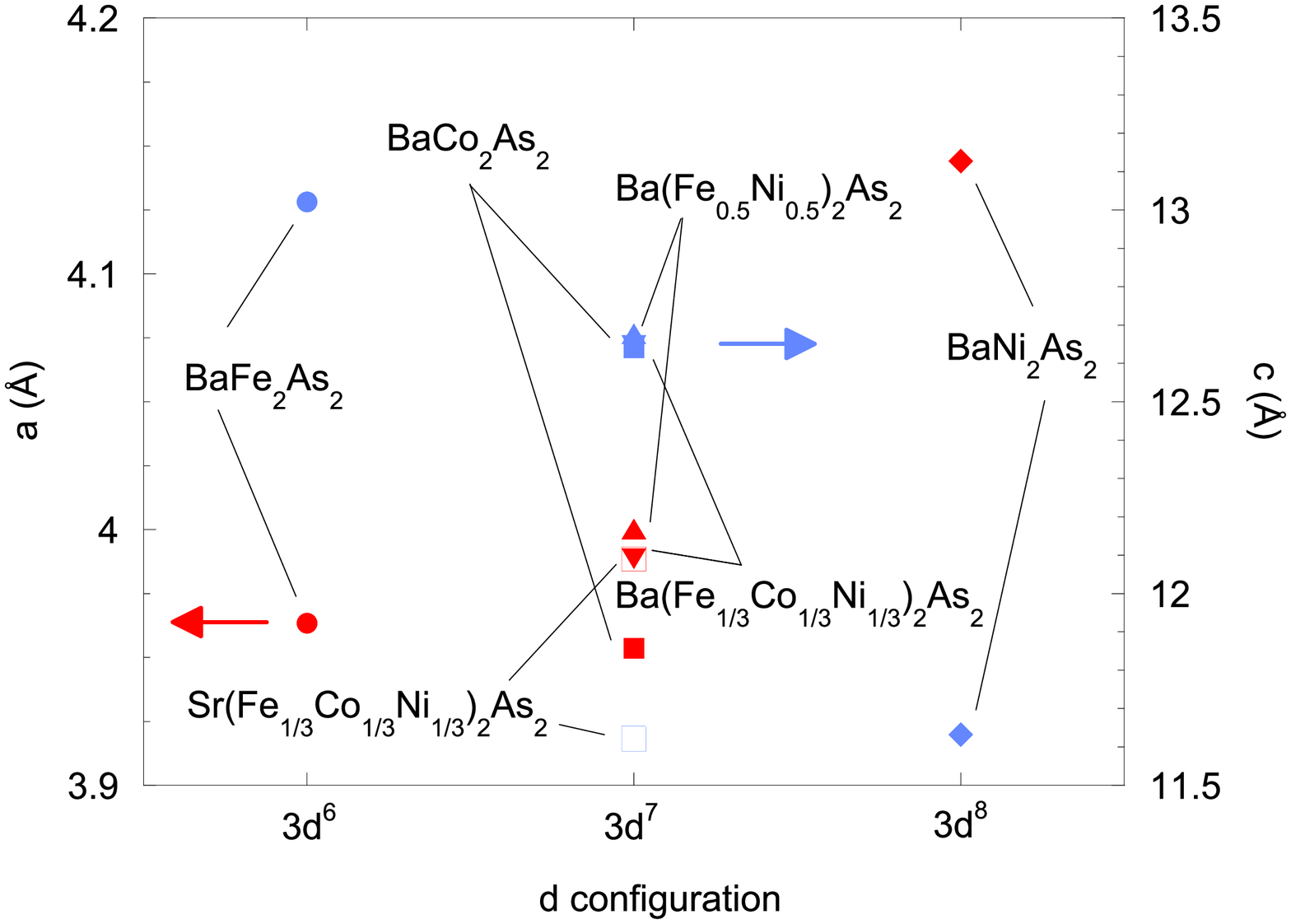}
\end{center}
\caption{{\bf Lattice constant of A$TM_{2}$As$_{2}$ (A = Ba, Sr, TM = Fe,Co,Ni).} Lattice parameters $a$ (red) and $c$ (blue) as a function of $3d$ configurations for BaFe$_{2}$As$_{2}$ ($\bullet$), BaCo$_{2}$As$_{2}$ ($\blacksquare$), Ba(Fe$_{1/3}$Co$_{1/3}$Ni$_{1/3}$)$_{2}$As$_{2}$ ($\blacktriangledown$), Ba(Fe$_{0.5}$Ni$_{0.5}$)$_{2}$As$_{2}$ ($\blacktriangle$), BaNi$_{2}$As$_{2}$ ($\blacklozenge$), and Sr(Fe$_{1/3}$Co$_{1/3}$Ni$_{1/3}$)$_{2}$As$_{2}$ ($\square$).}
\end{figure}

\begin{figure}[tbh]
  \begin{center}
    \includegraphics[width=9cm]{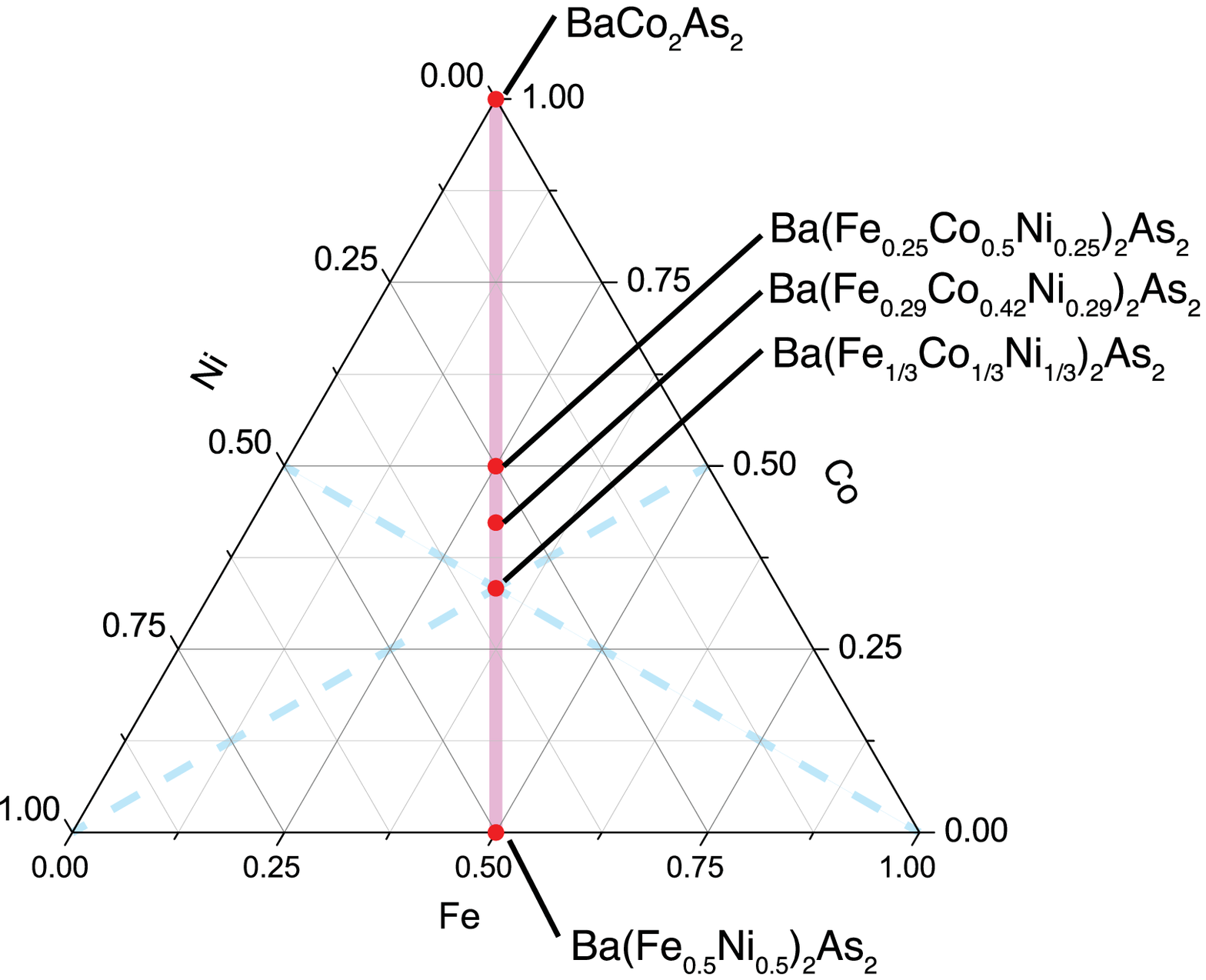}
  \end{center}
\caption{{\bf Ternary phase diagram}. A ternary phase diagram for Ba(Fe,Co,Ni)$_{2}$As$_{2}$. The red circles indicating the locations of BaCo$_{2}$As$_{2}$, Ba(Fe$_{0.25}$Co$_{0.5}$Ni$_{0.25}$)$_{2}$As$_{2}$, Ba(Fe$_{0.29}$Co$_{0.42}$Ni$_{0.29}$)$_{2}$As$_{2}$, {\23}, and Ba(Fe$_{0.5}$Ni$_{0.5}$)$_{2}$As$_{2}$, determined by energy dispersion spectroscopy.}
\end{figure}

\section{Single crystal refinements}
Single-crystal x-ray diffraction was performed at 150 K and 250 K for {\23} and at 250 K for {\sr} with Bruker APEX-II CCD system equipped with a graphite monochromator and a MoKa sealed tube ($\lambda$ = 0.71073 \AA). The crystallographic data obtained from refinements for {\23} and {\sr} are summarized in Table 1. Note that the final indices of the refinements $R_{1}$ are 1.01 \% (Ba at 250 K), 1.12 \% (Ba at 150 K), and 1.58 \% (Sr at 250 K), close to the best values for Ba 122 crystals \cite{kirsh12}, indicative of the high quality samples in which the doped transition metals are randomly distributed and do not form the clusters.

\begin{table}[htbp]
  \begin{center}
  \caption{Crystallographic data for $A$(Fe$_{1/3}$Co$_{1/3}$Ni$_{1/3}$)$_{2}$As$_{2}$ ($A$ = Ba, Sr) determined by single-crystal x-ray diffraction.}
  \begin{tabular}{llll}\hline\hline
    &Ba&Ba&Sr\\\hline
    Temperature&250~K&150~K&250~K\\
    Structure&tetragonal &tetragonal&tetragonal\\
    Space group&I4/mmm&I4/mmm&I4/mmm\\
    $a$ (\AA)&3.9920(3)&3.9826(3)&3.9885(8)\\
    $c$ (\AA)&12.6191(8)&12.6269(10)&11.621(5)\\
    $V$ (\AA$^{3}$)&201.10(3)&200.28(3)&184.87(9)\\
    $Z$ (formula unit/unit cell)&2&2&2\\
    $R_{1}$ ($I\geq 2\sigma (I)$)& 0.0101&   0.0112  &0.0158\\
    $wR_{2}$  (all data)&  0.0251& 0.0264 & 0.0344 \\
    Atomic coordinates (Wyckoff):&&&\\
    Ba/Sr (2a)&0, 0, 0&0, 0, 0&0, 0, 0\\
    Fe/Co/Ni (4d)&0.5, 0, 0.25&0.5, 0, 0.25&0.5, 0, 0.25\\           
    As (4e)&0.5, 0.5, 0.35160(3)&0.5, 0.5, 0.35171(3)&0.5, 0.5, 0.35840(7)\\
    Isotropic displacement &&&\\
    parameters $U_{eq}$ (\AA$^2$):&&&\\
    Ba/Sr&0.01058(8) &0.00740(8) &0.0122(2)\\
    Fe/Co/Ni&0.00961(10) & 0.00681(10)&0.0133(2)\\
    As&0.00913(9)& 0.00649(9)&0.01224(19)\\
    Bond lengths (\AA): &&&\\
    Ba/Sr-As&3.3875(3)& 3.3818(3)&3.2653(7)\\
    Fe/Co/Ni-As&2.3723(2)& 2.3695(2)&2.3588(6)\\
    Fe/Co/Ni-Fe/Co/Ni&2.8228(2)& 2.8161(2)&2.8203(4)\\
    Bond angles (deg):&&&\\
    As-Fe/Co/Ni-As &114.575(16)& 114.362(15)&115.44(4)\\
    As-Fe/Co/Ni-As &106.981(7)& 107.083(7)&106.57(2)\\
    \hline\hline
  \end{tabular}
  \end{center}
\end{table}

\section{Ternary phase diagram and possible competing phase}
Figure S2 shows a ternary phase diagram for Ba(Fe,Co,Ni)$_{2}$As$_{2}$ with red circles indicating the locations of BaCo$_{2}$As$_{2}$, Ba(Fe$_{0.25}$Co$_{0.5}$Ni$_{0.25}$)$_{2}$As$_{2}$, Ba(Fe$_{0.29}$Co$_{0.42}$Ni$_{0.29}$)$_{2}$As$_{2}$, {\23}, and Ba(Fe$_{0.5}$Ni$_{0.5}$)$_{2}$As$_{2}$,. Determined by energy dispersion spectroscopy, the compositions of Fe, Co, and Ni allow the samples to hold $3d^{7}$ configuration. Interestingly, the very-low-temperature charge transport for Ba(Fe$_{0.25}$Co$_{0.5}$Ni$_{0.25}$)$_{2}$As$_{2}$ reveals a resistive kink at 400 mK, possibly associated with a  phase transition (fig.~S3a), robust against magnetic fields, while that for Ba(Fe$_{0.25}$Co$_{0.5}$Ni$_{0.25}$)$_{2}$As$_{2}$ shows linear $T$ behavior, suppressed with applying field (fig.~S3b).

\begin{figure}[t]
  \begin{center}
    \includegraphics[width=9cm]{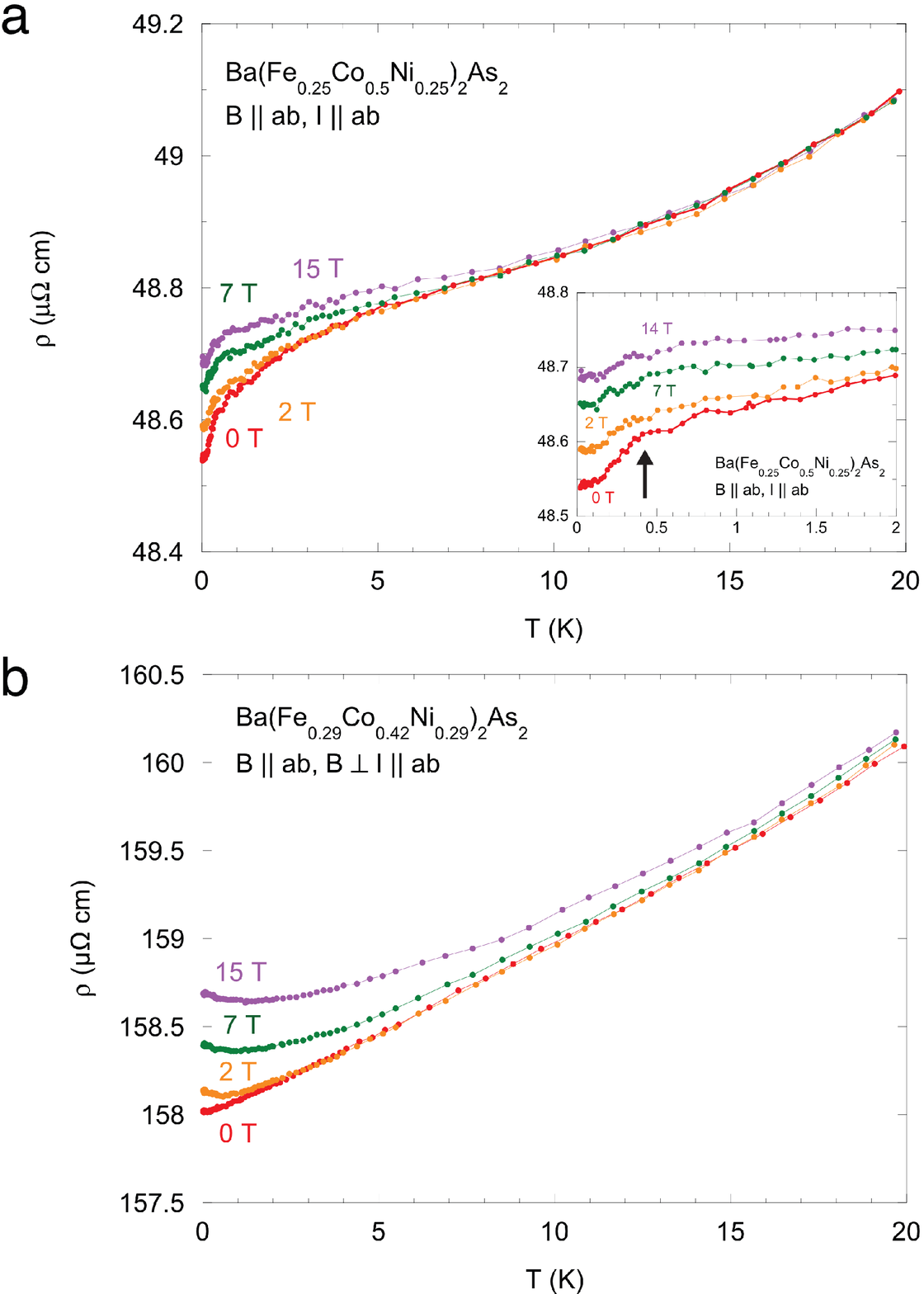}
  \end{center}
\caption{{\bf Resistivity for Ba(Fe,Co,Ni)$_{2}$As$_{2}$}. {\bf a}, temperature dependence of resistivity for Ba(Fe$_{0.25}$Co$_{0.5}$Ni$_{0.25}$)$_{2}$As$_{2}$. A resistive kink possibly involving a phase transition is observed at 400 mK. The transition temperature is robust against applying field. {\bf b}, temperature dependence of resistivity for Ba(Fe$_{0.29}$Co$_{0.42}$Ni$_{0.29}$)$_{2}$As$_{2}$, showing non-Fermi liquid behavior at zero field, diminished with magnetic field.}
\end{figure}

\begin{figure}[tbh]
\begin{center}
  \includegraphics[width=9cm]{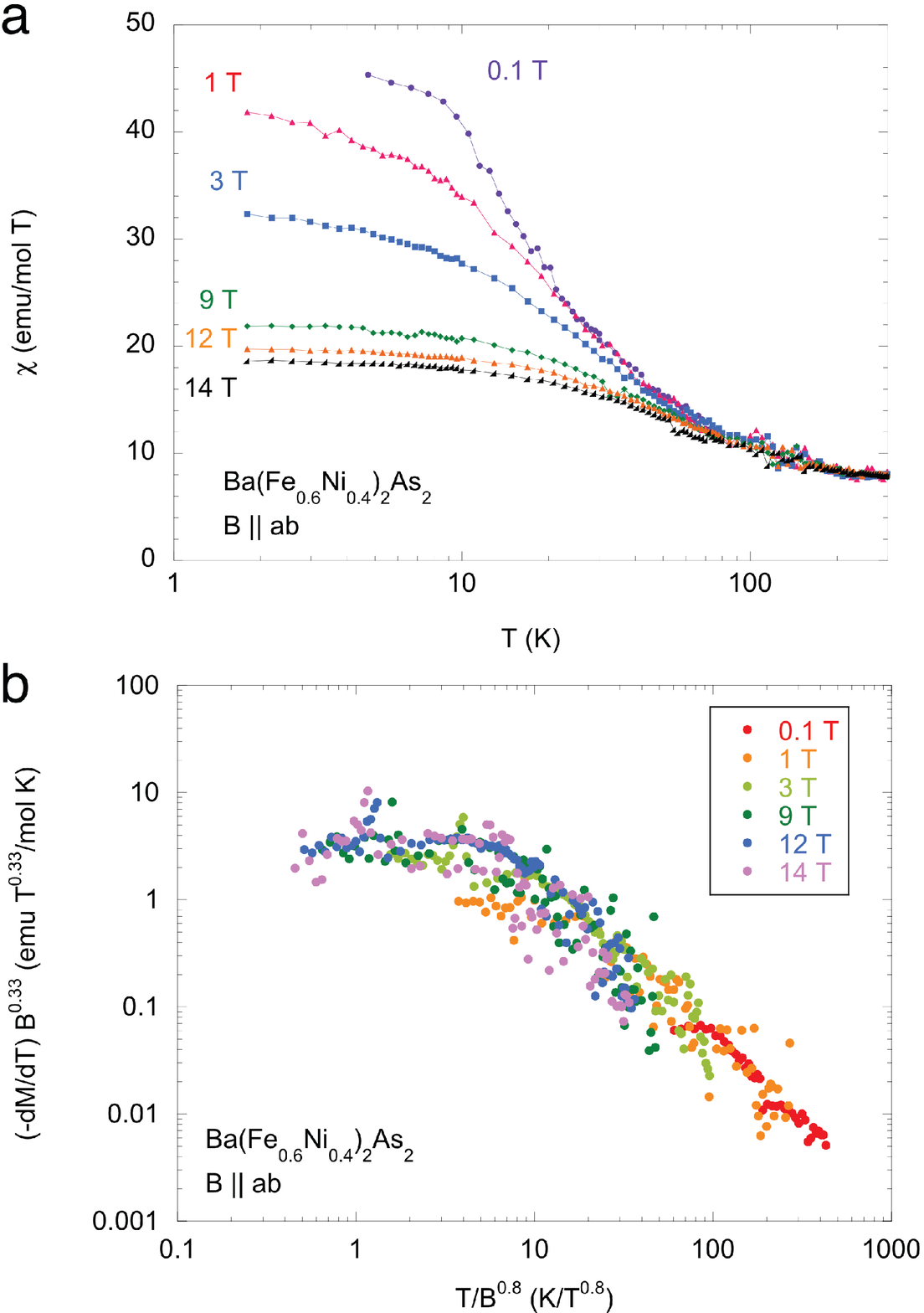}
\end{center}
\caption{{\bf Quantum critical scaling in the magnetization for Ba(Fe,Ni)$_{2}$As$_{2}$}. {\bf a}, Temperature dependence of the susceptibility $\chi$ for {Ba(Fe,Ni)$_{2}$As$_{2}$} ($B\parallel ab$). Less diverging behavior in $\chi$ implies the system is located slightly far from the quantum critical point. {\bf b}, Quantum critical scaling in the magnetization with critical exponents of $d/z=2/3$ and $z/y_{b} = 0.8$, slightly different from those extracted in {\23}.}
\end{figure}

Heavily electron doped Ba(Fe,Ni)$_{2}$As$_{2}$, assumedly sharing the same $3d^{7}$ configuration with {\Co} and {\23}, also shows non-Fermi liquid behavior in the magnetic susceptibility. As shown in fig.~S4a, the susceptibility divergetly increases with decreasing temperatures, followed by the saturation below 10 K even at $B=0$ T. This saturation at finite temperatures implies Ba(Fe,Ni)$_{2}$As$_{2}$ is located slightly away from a QCP. The non-Fermi liquid temperature dependence is strongly suppressed with applying magnetic field, indicative of the recovery of Fermi liquid regime at the applied field of 14 T. Similar to {\23}, the crossover from Fermi liquid to non-Fermi liquid indeed allows the quantum critical scaling in the magnetization with the critical exponents of $d/z=2/3$ and $z/y_{b} =0.8$, while the obtained $z/y_{b}$ is slightly different from that for {\23}.

\section{Pressure measurements}
A non-magnetic piston-cylinder pressure cell was used for transport measurements under pressure up to 1.99 GPa, using a 1 : 1 ratio of n-Pentane to 1-methyl-3-butanol as the pressure medium and superconducting temperature of lead as pressure gauge at base temperature. All transport measurements were performed on the same Ba(Fe$_{1/3}$Co$_{1/3}$Ni$_{1/3}$)$_{2}$As$_2$ crystal with 200 $\mu$m thickness using four point contacts made with silver epoxy. The pressure and temperature dependence of the resistivity were measured during warming process in a Quantum Design PPMS.

\section{Non-Fermi liquid to Fermi liquid crossover in the resistivity}
The Non-Fermi liquid behavior in the temperature dependence of resistivity is strongly suppressed with magnetic field. Upon applying magnetic field, the recovery of Fermi liquid behavior, $\rho\propto T^{2}$, is observed, independent of applied magnetic field directions at low temperatures (fig.S5a and b). The crossover temperature from non-Fermi liquid to Fermi liquid behavior, $T_{FL}$, is extracted from the deviation from $T^{2}$-fit.

\begin{figure}[thb]
\includegraphics[width=15cm]{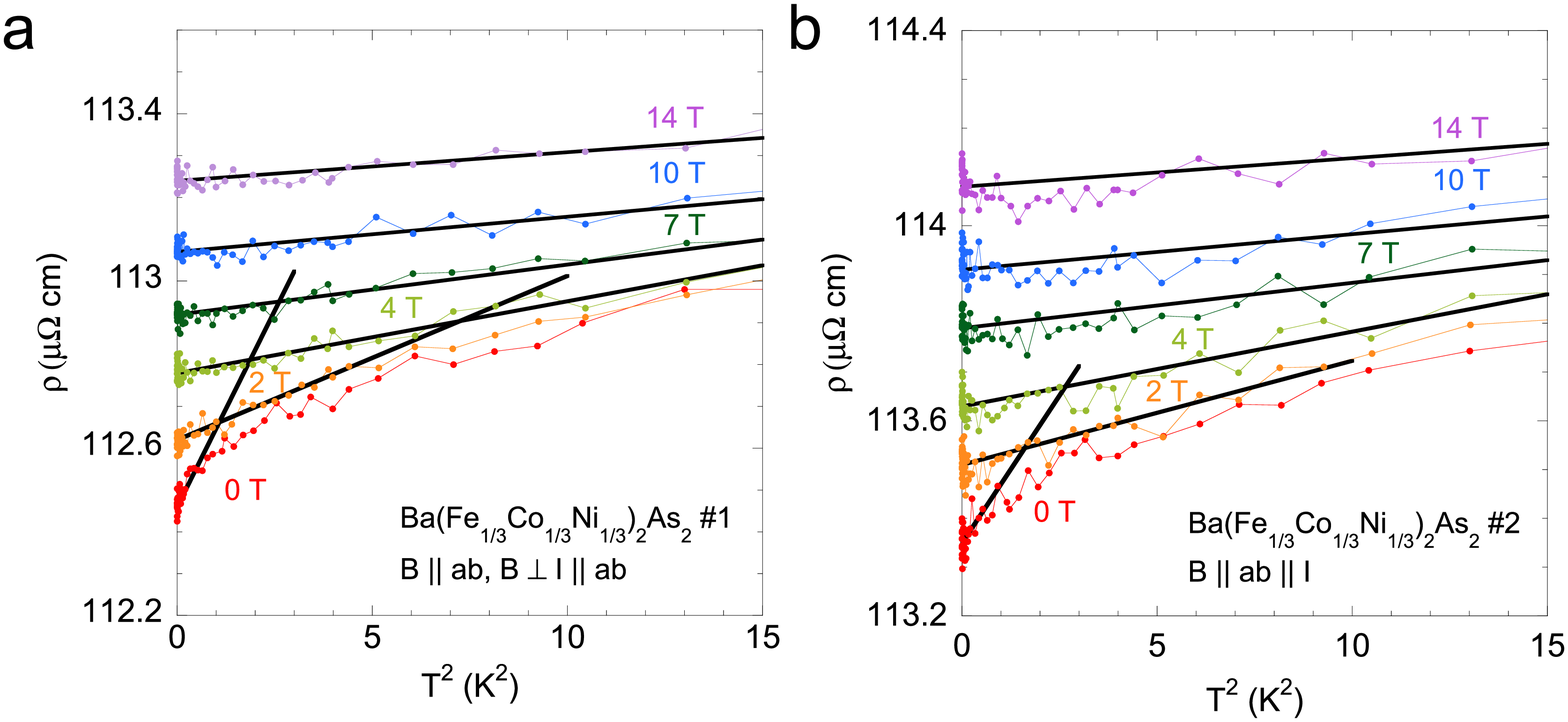}
\caption{{\bf Non-Fermi liquid to Fermi liquid crossover in the resistivity for BaCo$_{2}$As$_{2}$}. {\bf a}, Resistivity as a function of $T^{2}$ for sample \#1 in the configuration of $B\parallel  ab, B \perp I \parallel ab$ and {\bf b}, for sample \#1 in the configuration of $B\parallel  ab \parallel I$. Solid lines are linear fits to the data using $\rho = \rho_{0} + AT^{2}.$ }
\end{figure}

\section{Quantum critical ferromagnetic scatterings in $\mathrm{{\bf BaCo_{2}As_{2}}}$}
As evinced by the observation of the enhanced Wilson ratio and violation of the Koringa ratio, {\Co} is located close to the ferromagnetic quantum instabilities. The instabilities actually cause unusual scatterings in the charge transport for {\Co} (fig.~S6a). Unlike {\23}, the temperature dependence of resistivity for {\Co} is not sublinear, but superlinear. To clarify the exponent of the temperature dependence, we plot the resistivity as a function of $T^{2}$ (fig.~S6b), expected for Fermi liquid, and of $T^{5/3}$ (fig.~S6c), expected for three dimensional quantum critical ferromagnets. Very similar to quantum critical ferromagnetic metal ZrZn$_{2}$ \cite{suthe12}, the perfect linear-in-$T^{5/3}$ dependence of the resistivity below $T=20$ K highlights the presence of abundant quantum critical scatterings in {\Co}, robust against applied field, even up to 15 T.

\begin{figure}[thb]
  \begin{center}
    \includegraphics[width=9cm]{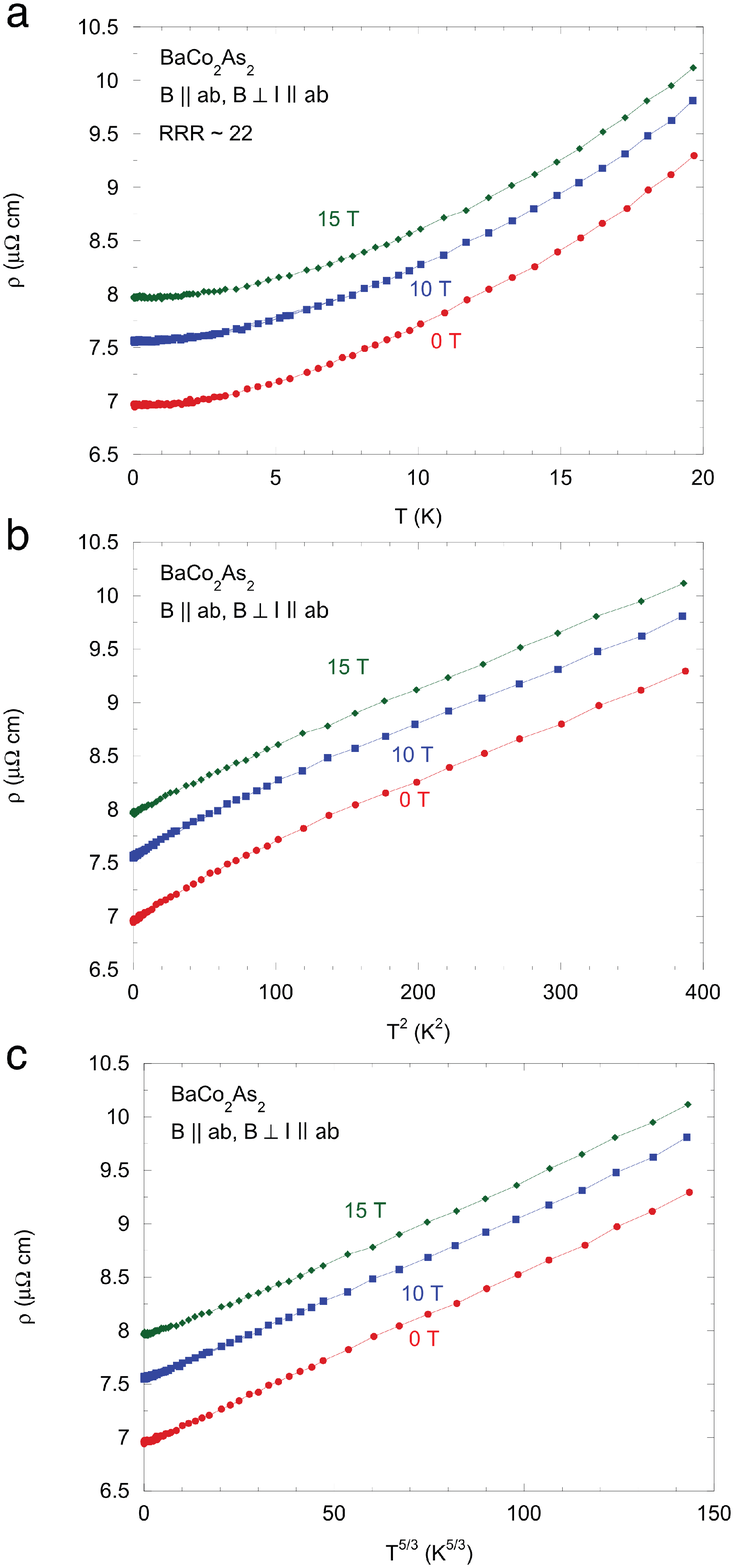}
  \end{center}
\caption{{\bf Ferromagnetic quantum critical scattering for BaCo$_{2}$As$_{2}$}. {\bf a}, Temperature dependence of resistivity for {\Co}. The resistivity for {\Co} as a function of {\bf a}, $T^{2}$ and  {\bf b}, $T^{5/3}$, suggestive of ferromagnetic quantum critical scattering in 3D systems and reminiscent of marginal Fermi liquid ZrZn$_{2}$.}
\end{figure}

\section{Isotropy of non-Fermi liquid behavior and ${\bf \Gamma(T,B)}$ scaling}
\begin{figure}[tbh]
\includegraphics[width=15cm]{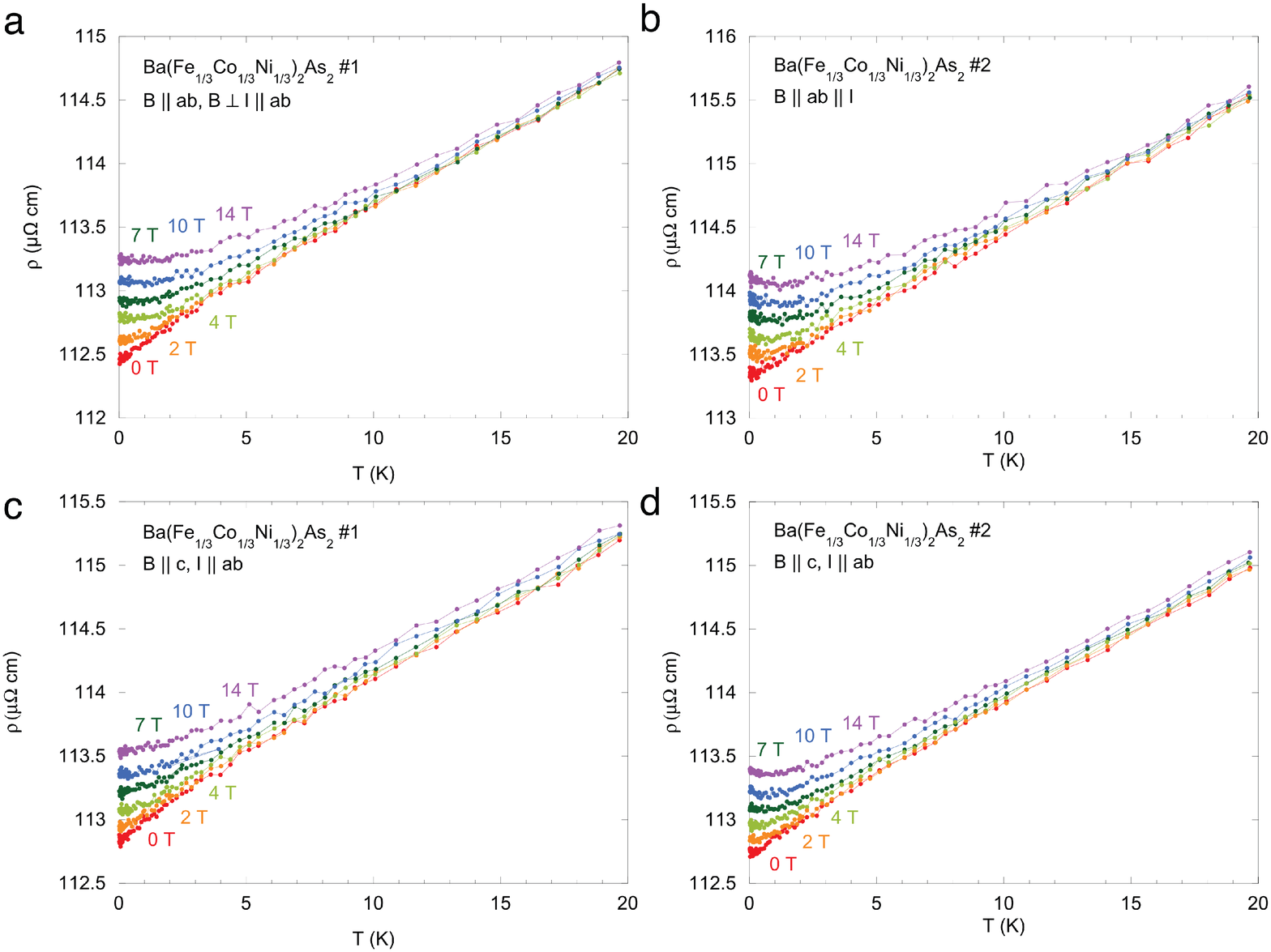}
\caption{{\bf Isotropic magnetoresistance for {\23}}. Resistivity as a function of $T$ for {\bf a}, sample \#1 with $B\parallel ab$ and  $B\perp I\parallel ab$, {\bf b}, sample \#2 with $B\parallel I\parallel ab$, {\bf c}, sample \#1 with $B\parallel c$ and  $BI\parallel ab$, and {\bf d}, sample \#2 with $B\parallel c$ and $BI\parallel ab$.}
\end{figure}

\begin{figure}[tbh]
\includegraphics[width=15cm]{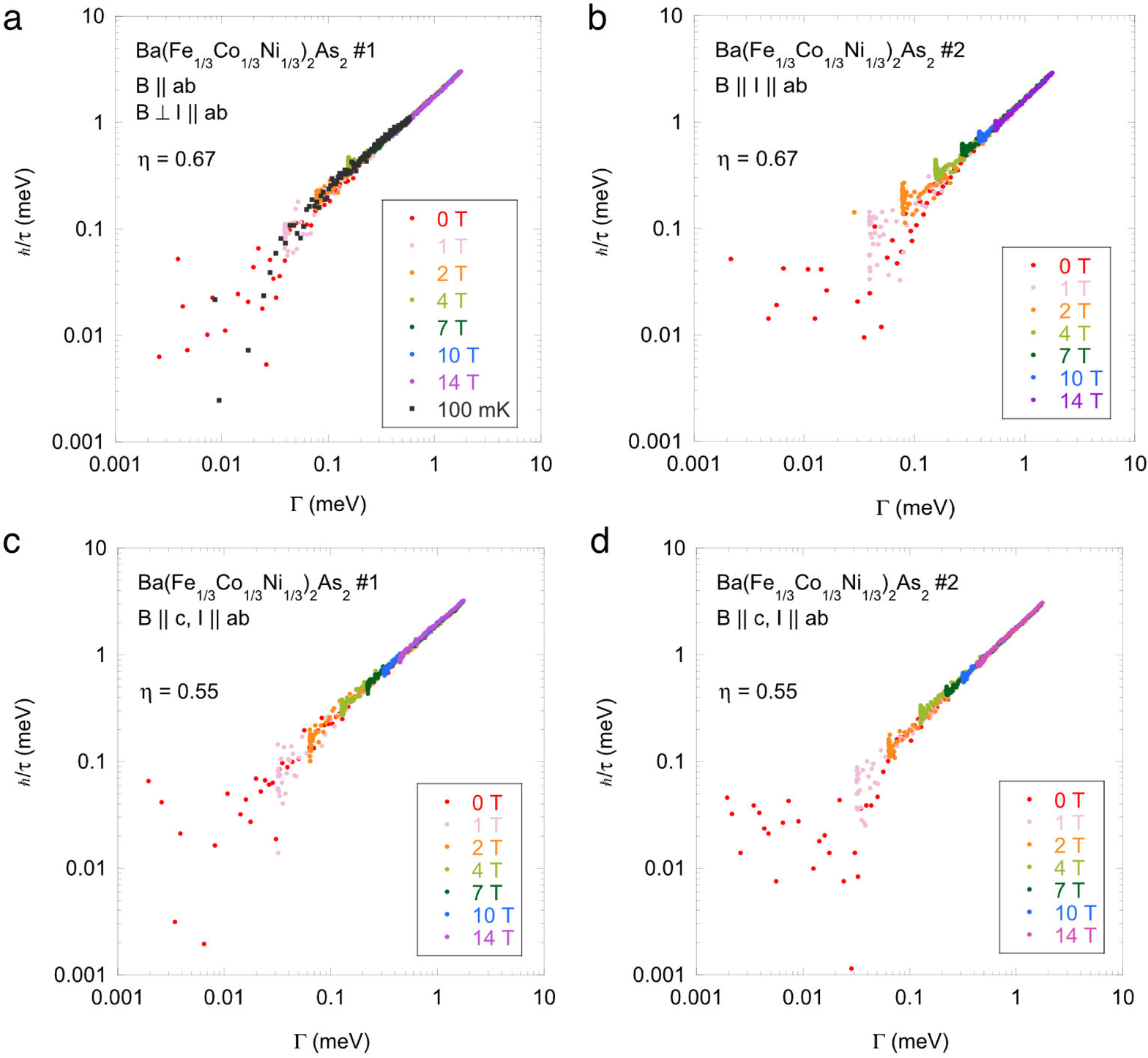}
\caption{{\bf Anisotropy of $\Gamma$(T,B) scaling for {\23} with different field orientations}. Resistivity as a function of $\Gamma\equiv \sqrt{(k_{B}T)^{2}+(\eta \mu_{B}B)^{2}}$ for {\bf a}, sample \#1 with $B\parallel ab$ and  $B\perp I\parallel ab$ together with sample \#4 (scaled), {\bf b}, sample \#2 with $B\parallel I\parallel ab$, {\bf c}, sample \#1 together with sample \#4 (scaled) and {\bf d}, \#2 with $B\parallel c$ and  $I\parallel ab$.}
\end{figure}

Despite of the quasi layered structure, the non-Fermi-liquid magnetoresistance of {\23} is independent of applied field orientations. Figure S6 shows the temperature dependence of resistivity in different applied field configurations. Independent of the applied field orientations, the quasi-$T$-linear dependence of resistivity at zero field is suppressed with field, suggesting the spatial dimensionality is three (fig.1e in the main text).

Obtaining from the magnetoresistance as shown in fig.~S7, we plot $\Gamma (T,B)$ scaling in the resistivity, independent of field directions with respect to the current direction. For the in-plane field orientations ($B\parallel ab$), either longitudinal ($B\parallel I$) or transverse magnetoresistance ($B\perp I$) provides the ratio of scaling parameters of $\eta=0.55$ (fig.~S8a and b). By contrast, magnetoresistance in the perpendicular field orientation ($B\parallel c$) gives the ratio of $\eta =0.67$ (fig.~S8 c and d). The anisotropy of the scaling parameter ratio $\eta_{B\parallel ab}/\eta_{B\parallel c}$ is $0.82$ and close to unity, suggesting isotropic scatterings. The non-Fermi liquid behavior in the magnetization measurements is also independent of applied field orientations, evidenced by the isotropy between the magnetization along $B\parallel c$ and $B\parallel ab$ (fig.~S9). The critical exponent of magnetization $\delta$ is 1.34, obtained by a fit to the data using $M\sim B^{\frac{1}{\delta}}$.

\clearpage
\begin{figure}[tbh]
\begin{center}
  \includegraphics[width=9cm]{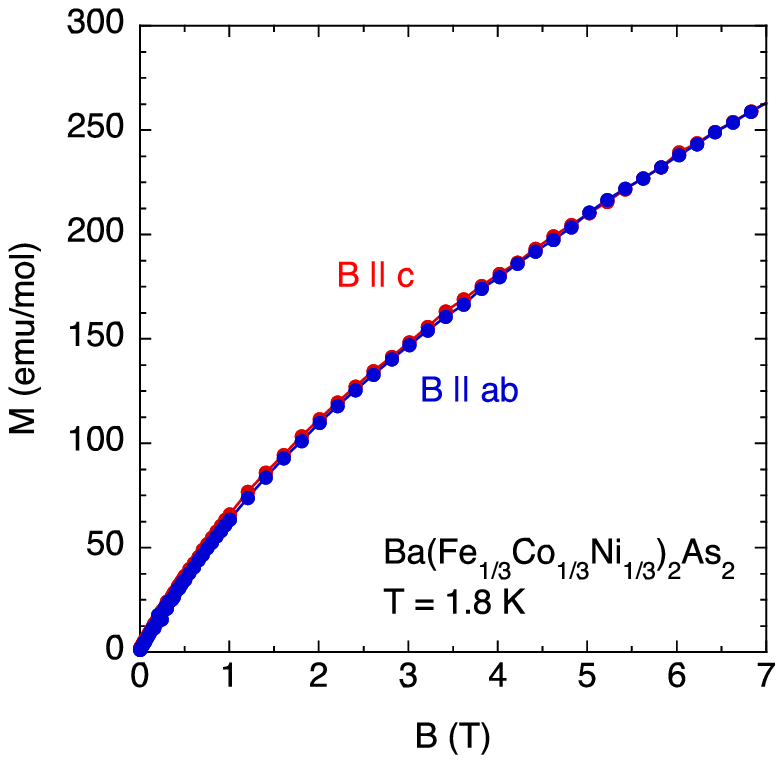}
\end{center}
\caption{{\bf Isotropic magnetization for Ba(Fe,Co,Ni)$_{2}$As$_{2}$}. Field dependence of magnetization along $B\parallel ab$ and $B\parallel c$ at $T$ = 1.8 K. Despite of the layered structure, anisotropy of magnetization is negligible, suggesting the system is three dimensional. Inset: magnetization versus B ($\parallel ab$) at 1.8 K. A red line is a fit to the data using $M\sim B^{\frac{1}{\delta}}$ with $\delta =1.34$.}
\end{figure}

\section{Quantum critical scaling}
The quantum critical scaling observed in magnetization (fig.~2c) and specific heat (fig.~2d) implies the presence of universal function of $T/B$ in the free energy. We can assume the generic form for the free energy $F$ as,
\begin{equation}
  F(B,T)=T^{\frac{d+z}{z}}\tilde{f}_{F}\left (\frac{B}{T^{y_{b}/z}}\right)=B^{\frac{d+z}{y_{b}}}f_{F}\left (\frac{T}{B^{z/y_{b}}}\right),
\end{equation}
where $y_{b}$ is the scaling exponent related to magnetic field $B$, $d$ is the spatial dimension, and $z$ is the dynamical exponent. Assuming this form of free energy, we can derive magnetization $M=\partial F/\partial B$ and specific heat $C/T = -\partial^{2}F/\partial T^{2}$. The magnetization is written by,
\begin{equation}
  M=B^{(d+z)/y_{b}-1}f_{M}\left(\frac{T}{B^{z/y_{b}}}\right),
\end{equation}
where the scaling function $f_{M}$ is also a universal function of $x=T/B^{z/y_{b}}$, given by,
\begin{equation}
  f_{M}(x) = [(d+z)/y_{b}]f_{F}(x)-(z/y_{b})xf_{F}^{\prime}(x). 
\end{equation}
To extract the critical exponents, we obtain the derivative of $M$,
\begin{equation}
  -\frac{dM}{dT}=B^{d/y_{b}-1}f_{M}^{\prime}\left(\frac{T}{B^{z/y_{b}}}\right).
\end{equation}
By comparing this with the scaling relation observed in fig.~2c, the critical exponents yield,
\begin{equation}
  \left\{
      \begin{array}{ccc}
        d/y_{b}-1&=&-1/3\\
        z/y_{b}&=&1.
      \end{array}
      \right .
\end{equation}
These equations provide
\begin{equation}
  \left\{
      \begin{array}{ccc}
        z&=&y_{b}\\
        d/z&=&2/3.
      \end{array}
      \right .
\end{equation}
Likewise, the specific heat can be given by,
\begin{equation}
  \frac{C(B,T)}{T}=-\frac{\partial^{2}F}{\partial T^{2}}=T^{(d-z)/z}\tilde{f}_{C}\left(\frac{B}{T^{y_{b}/z}}\right),
\end{equation}
where $\tilde{f}_{C}(\tilde{x})$ is a scaling function of $\tilde{x}=B/T^{y_{b}/z}$,
\begin{eqnarray}
  \tilde{f}_{C}(\tilde{x}) & = & (d(d+z)/z^{2})\tilde{f}_{F}(\tilde{x})-(y_{b}(2d+z-y_{b})/z^{2})\tilde{x}\tilde{f}_{F}^{\prime}(\tilde{x})+(y_{b}^{2}/z^{2})\tilde{x}^{2}\tilde{f}_{F}^{\prime\prime}(\tilde{x})\\
  & = & (d(d+z)/z^{2})\tilde{f}_{F}(0) + \tilde{g}_{C}(\tilde{x}),
\end{eqnarray}
where, $\tilde{g}_{C}(\tilde{x})$ is field-dependent part of $\tilde{f}_{C}(\tilde{x})$. Using this expression, we can extract field dependent part of specific heat,
\begin{equation}
  \frac{\Delta C_{e}(B,T)}{T} =  \frac{\Delta C_{e}(B,T)}{T} - \frac{\Delta C_{e}(0,T)}{T} = T^{\frac{d-z}{z}}\tilde{g}_{C}(B/T^{y_{b}/z}) = B^{\frac{d-z}{y_{b}}}g_{C}(T/B^{z/y_{b}}),  
\end{equation}
where $g_{C}(x)$ is temperature-dependent part of $f_{C}(x)$. By comparing this with the scaling relation in fig.~2d, we obtain the critical exponents yielding,
\begin{equation}
  \left\{
      \begin{array}{ccc}
        (d-z)/y_{b}&=&-1/3\\
        z/y_{b}&=&1,
      \end{array}
      \right .
\end{equation}
also providing the same parameters as the eqs.~(S6), namely,
\begin{equation}
  \left\{
      \begin{array}{ccc}
        z&=&y_{b}\\
        d/z&=&2/3.
      \end{array}
      \right .
\end{equation}

\section{Scaling function and Fermi to non-Fermi liquid crossover}
The obtained scaling relations clearly show the Fermi to non-Fermi liquid crossover behavior. For $T/B \gg 1$, we observe non-Fermi liquid diverging behavior in the susceptibility, $\chi \propto T^{-1/3}$, implying $f_{M}(x)\propto x^{-1/3} $. On the other hand, in the other limit of $T/B \ll 1$, we observed temperature independent susceptibility, suggestive of the recovery of FL regime. From these observations, we can write the asymptotic forms of $f_{M}(x)$, 
\begin{equation}
  f_{M}(x)\propto\left\{
      \begin{array}{lll}
        x^{-1/3}& T \gg B & \mathrm{quantum~critical~regime}\\
        const + O(x^{2}) & T \ll B& \mathrm{Fermi~liquid~regime}.
      \end{array}
      \right .
\end{equation}
These asymptotic forms allow us to specify a universal function,
\begin{equation}
  f_{M}(x)=c(x^{2}+a^{2})^{-1/6},
\end{equation}
reproducing the behavior in $x\ll 1$ and $x\gg 1$ limits. Using eq.~(S6),
\begin{equation}
  M = c B^{2/3}(x^{2}+a^{2})^{-1/6}.
\end{equation}
The peak position in $dM/dT$ gives the crossover temperature $T^{\ast}$ by using $\frac{d}{dT}(dM/dT)$ = 0, which gives,
\begin{equation}
  T^{\ast}/B = x^{\ast} = \sqrt{3}a/2.
\end{equation}
Extracted from this equation, $T^{\ast}(B)$ is plotted in the phase diagram (fig.~3c).

Similarly, $T^{\ast}$ can also be extracted from the scaling in the specific heat, which follows the Maxwell relation linking the entropy to the magnetization,
\begin{equation}
  \frac{\partial S(B,T)}{\partial B} = \frac{\partial M(B,T)}{\partial T}.
\end{equation}
Integrating both sides with respect to B, we can obtain,
\begin{equation}
  \int^{B}_{0}\frac{\partial S(B,T)}{\partial B} dB =  \int^{B}_{0}\frac{\partial M(B,T)}{\partial T} dB.
\end{equation}
Since
\begin{equation}
  \int^{B}_{0}\frac{\partial S(B,T)}{\partial B} dB = S(B,T)-S(0,T) = \int^{T}_{0}\frac{\Delta C_{e}(B,T)}{T} dT,
\end{equation}
using eq.(S6), (S18), and (S19), we get,
\begin{equation}
  \frac{\Delta C_{e}(B,T)}{T}= \frac{\partial^{2}}{\partial T^{2}}\int^{B}_{0}M(B,T)dB = \int^{B}_{0}B^{-4/3}f^{\prime\prime}_{M}(x)dB,
\end{equation}
where
\begin{equation}
  f^{\prime\prime}_{M}(x) = -\frac{c}{3}\left (x^{2}+a^{2}\right )^{-7/6}\left [ 1-\frac{7}{3}\frac{x^{2}}{x^{2}+a^{2}}\right].
\end{equation}
The peak positions in the scaling function of $\Delta C_{e}/T$ obtained from a fit to the data give the crossover temperature $T^{\ast}(B)$, consistent with $T^{\ast}$ from $M$ as plotted in the phase diagram (fig.~3c).

\section{Absence of Mooij correlations and validity of the Matthiessen's rule}
Transport properties in highly disordered metals show strong deviations from those described by the Boltzmann model. In the disordered metals, different scattering processes can no longer be treated independently, in other words, Matthiessen's rule breaks down. With introducing disorders, the residual resistivity in conventional metals increases toward the Mott-Ioofe-Regel limit, leading to a change of sign of the temperature coefficient of resistivity $d\rho/dT$ from positive to negative at high temperatures \cite{lee85}. The sign change anticorrelates with the residual resistivity, known as Mooij correlations \cite{ciuch18}. In the Mooij regime, polaronic renormalization of disorder plays an important role in the scattering mechanism, causing the breakdown of Matthiessen's rule.

On the other hand, in {\23}, which can be considered as a highly disordered version of {\Co}, the introduction of disorder by counter-doping to Co sites enhances the residual resistivity, but causes no decrease in the slope of resistivity at high temperatures, resulting in a simple parallel shift of the resistivity. This indicates the present system is not in the Mooij regime and allows us to extract inelastic scattering part using the Matthiessen's rule.

\begin{figure}[tbh]
  \begin{center}
    \includegraphics[width=10cm]{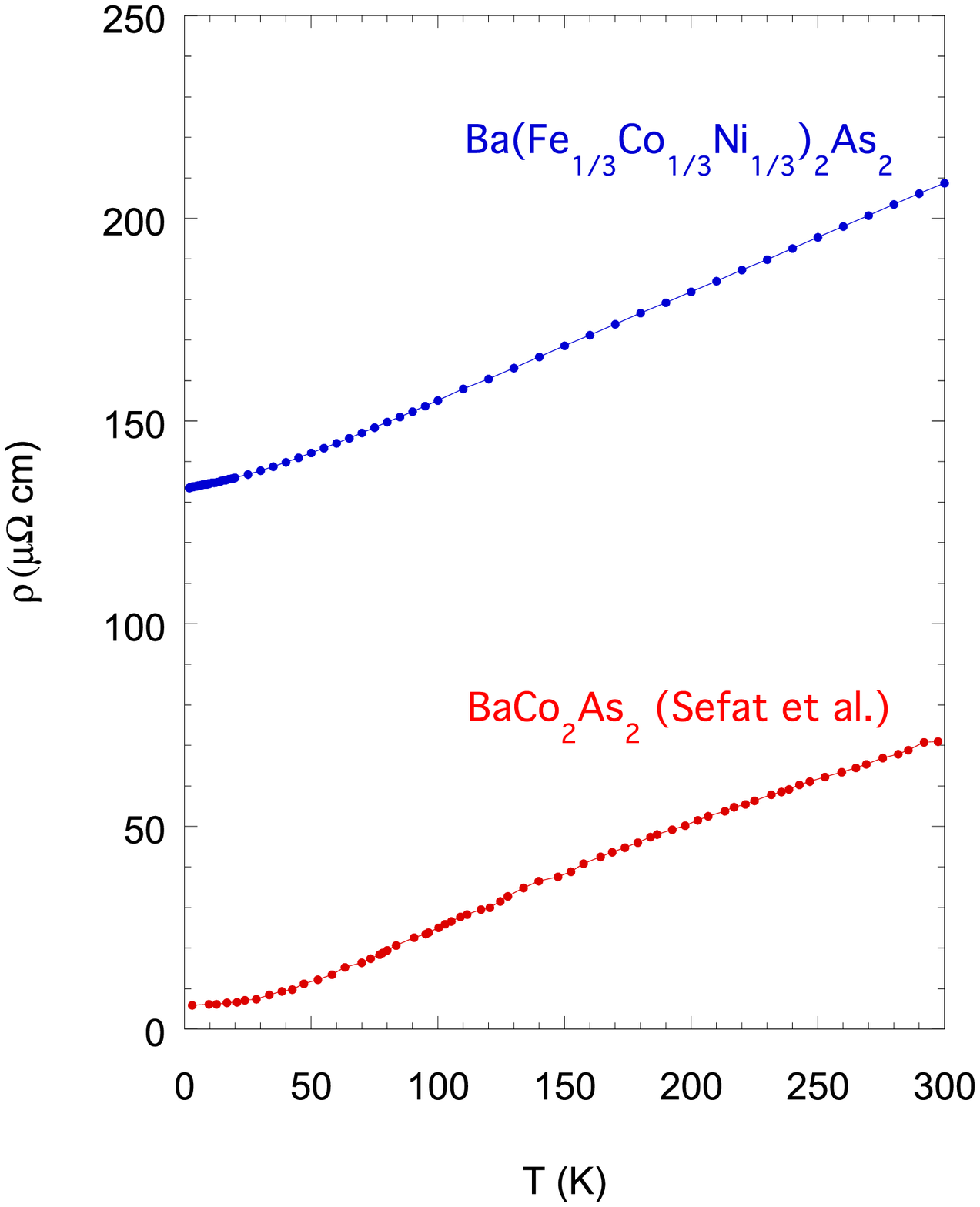}
    \end{center}
\caption{{\bf absence of Mooij correlations.} Temperature dependence of resistivity for clean {\Co} (taken from \cite{sefat09}) and highly disordered {\23}. Introduction of disorder causes no change of the slop of resistivity at high temperatures, inconsistent with the Mooij correlations.} 
\end{figure}

\end{document}